\documentclass[aoas]{imsart}
\usepackage{newtxtext}

\RequirePackage{amsthm,amsmath,amsfonts,amssymb}
\RequirePackage[authoryear, sectionbib]{natbib}%% uncomment this for author-year citations
\RequirePackage[colorlinks,citecolor=blue,urlcolor=blue]{hyperref}
\RequirePackage{graphicx}

\usepackage{booktabs} 
\usepackage{fix-cm}
\usepackage{chapterbib} % Enables section-specific bibliography

\begin{document}

\startlocaldefs
%%%%%%%%%%%%%%%%%%%%%%%%%%%%%%%%%%%%%%%%%%%%%%
%%                                          %%
%% Uncomment next line to change            %%
%% the type of equation numbering           %%
%%                                          %%
%%%%%%%%%%%%%%%%%%%%%%%%%%%%%%%%%%%%%%%%%%%%%%
%\numberwithin{equation}{section}
%%%%%%%%%%%%%%%%%%%%%%%%%%%%%%%%%%%%%%%%%%%%%%
%%                                          %%
%% For Axiom, Claim, Corollary, Hypothesis, %%
%% Lemma, Theorem, Proposition              %%
%% use \theoremstyle{plain}                 %%
%%                                          %%
%%%%%%%%%%%%%%%%%%%%%%%%%%%%%%%%%%%%%%%%%%%%%%
\theoremstyle{plain}
\newtheorem{axiom}{Axiom}
\newtheorem{claim}[axiom]{Claim}
\newtheorem{theorem}{Theorem}[section]
\newtheorem{lemma}[theorem]{Lemma}
\newtheorem{cor}[theorem]{Corollary}
\newtheorem{prop}[theorem]{Proposition}

%%%%%%%%%%%%%%%%%%%%%%%%%%%%%%%%%%%%%%%%%%%%%%
%%                                          %%
%% For Assumption, Definition, Example,     %%
%% Notation, Property, Remark, Fact         %%
%% use \theoremstyle{definition}            %%
%%                                          %%
%%%%%%%%%%%%%%%%%%%%%%%%%%%%%%%%%%%%%%%%%%%%%%
\theoremstyle{definition}
\newtheorem{definition}[theorem]{Definition}
\newtheorem{const}[theorem]{Construction}
\newtheorem{conj}[theorem]{Conjecture}
\newtheorem{rem}[theorem]{Remark}
\newtheorem*{example}{Example}
\newtheorem*{fact}{Fact}
%%%%%%%%%%%%%%%%%%%%%%%%%%%%%%%%%%%%%%%%%%%%%%
%% Please put your definitions here:        %%
%%%%%%%%%%%%%%%%%%%%%%%%%%%%%%%%%%%%%%%%%%%%%%

\newcommand{\sub}{\operatorname{sub}}
\newcommand{\quot}{\operatorname{quot}}
\newcommand{\bw}{\bigwedge}
\newcommand{\Avs}{\operatorname{Av}^{\operatorname{sign}}}
\newcommand{\bad}{\operatorname{bad}}
\newcommand{\sign}{\operatorname{sign}}
\newcommand{\id}{\operatorname{id}}
\newcommand{\defeq}{\vcentcolon=}
\newcommand{\eqdef}{=\vcentcolon}
%We can even define a new command for \newcommand!
\newcommand\nc{\newcommand}
\nc{\on}{\operatorname}
\nc\renc{\renewcommand}
\nc{\mycomment}[1]{}
\nc{\R}{\mathbb R}
\nc{\BC}{\mathbb C}
\nc{\BQ}{\mathbb Q}
\nc{\BZ}{\mathbb Z}
\nc{\BN}{\mathbb N}
\nc{\BS}{\mathbb S}
\nc{\BA}{\mathbb A}
\nc{\BP}{\mathbb P}
\nc{\tr}{\text{tr}}
\nc{\Hom}{\on{Hom}}
\nc{\wt}{\widetilde}
\nc{\vspan}{\on{span}}
\nc{\ord}{\on{ord}}
\nc{\im}{\on{im}}
\nc{\Mat}{\on{Mat}}
\nc{\can}{\on{can}}
\nc{\coker}{\on{coker}}
\nc{\ev}{\on{ev}}
\nc{\Tr}{\on{Tr}}
\nc{\End}{\on{End}}
\nc{\swap}{\on{swap}}
\nc{\Set}{\on{Set}}
\nc{\bC}{{\mathbf C}}
\nc{\bc}{{\mathbf c}}
\nc{\bD}{{\mathbf D}}
\nc{\bd}{{\mathbf d}}
\nc{\bE}{{\mathbf E}}
\nc{\be}{{\mathbf e}}
\nc{\bF}{{\mathbf F}}
\nc{\bff}{{\mathbf f}}
\nc{\CE}{\mathcal E}
\nc{\CO}{\mathcal O}
\nc{\CC}{\mathcal C}
\nc{\SC}{\mathscr C}
%\renc{\mod}{\on{-mod}} %Careful - turn this off in a number theory setting
\newcommand{\spec}{\text{spec}}
\nc{\adj}{\on{adj}}
\nc{\tensor}[3]{#1 \underset{#2}\otimes #3}
\nc{\Nat}{\on{Nat}}
\nc{\op}{\on{op}}
\nc{\Funct}{\on{Funct}}
\nc{\Ob}{\on{Ob}}
\nc{\fR}{\mathfrak{R}}
\nc{\Vect}{\on{Vect}}
\nc{\ns}{\on{non-spec}}
\nc{\ol}{\overline}
\nc{\univ}{\on{univ}}
\nc{\Maps}{\on{Maps}}
\nc{\bdd}{\on{bdd}}
\nc{\cont}{\on{cont}}
\nc{\Sym}{\on{Sym}}
\nc{\vol}{\on{vol}}
\nc{\supp}{\on{supp}}
\nc{\Lie}{\on{Lie}}
\nc{\master}{\on{master}}
\nc{\pt}{\on{pt}}
\nc{\funcon}{\on{Funct}_{\on{cont}}}
\nc{\funconpw}{\on{Funct}_{\on{cont},\on{pw-lin}}}
\nc{\ts}{\textsc}
\nc{\codim}{\on{codim}}
\nc{\fm}{\mathfrak m}
\nc{\SR}{\mathscr R}
\nc{\Tor}{\on{Tor}}
\nc{\Ext}{\on{Ext}}
\nc{\Syn}{\on{Syn}}

\endlocaldefs

\begin{frontmatter}
\title{Spectral-Stimulus Information for Self-Supervised
Stimulus Encoding}
%\title{A sample article title with some additional note\thanksref{t1}}
\runtitle{Spectral-Stimulus Information for Self-Supervised
Stimulus Encoding}
%\thankstext{T1}{A sample additional note to the title.}

\begin{aug}
%%%%%%%%%%%%%%%%%%%%%%%%%%%%%%%%%%%%%%%%%%%%%%%
%% Only one address is permitted per author. %%
%% Only division, organization and e-mail is %%
%% included in the address.                  %%
%% Additional information can be included in %%
\author[A,B]{\fnms{Jared}~\snm{Deighton} \ead[label=e1]{jared.deighton@simmons.edu}},
\author[C]{\fnms{Wyatt}~\snm{Mackey}\ead[label=e2]{wyatt.t.mackey.civ@army.mil}},
\author[C]{\fnms{Ioannis}~\snm{Schizas}\ead[label=e3]{ioannis.d.schizas.civ@army.mil}},
\author[C]{\fnms{David L.}~\snm{Boothe}\ead[label=e4]{david.l.boothe7.civ@army.mil}}
\and
\author[A]{\fnms{Vasileios}~\snm{Maroulas}\thanks{Corresponding author}\ead[label=e5]{vmaroula@utk.edu}}

%%%%%%%%%%%%%%%%%%%%%%%%%%%%%%%%%%%%%%%%%%%%%%
%% Addresses                                %%
%%%%%%%%%%%%%%%%%%%%%%%%%%%%%%%%%%%%%%%%%%%%%%
\address[A]{Department of Mathematics, The University of Tennessee, Knoxville, TN 37996, United States of America \printead[presep={,\ }]{e5}}

\address[B]{Department of Computer, Data, \& Mathematical Sciences, Simmons University, Boston, MA 02115, United States of America\printead[presep={,\ }]{e1}}

\address[C]{The U.S. Army Combat Capabilities Development Command Army Research Laboratory, Aberdeen Proving Ground, MD 21005, United States of America \printead[presep={,\ }]{e2,e3,e4}}

\end{aug}

\begin{abstract}
Mammalian spatial navigation relies on specialized neurons, such as place and grid cells, which encode position based on self-motion and environmental cues. While extensive research has explored the computational role of grid cells, the principles underlying efficient place cell coding remain less understood. Existing spatial information rate measures primarily assess single-neuron encoding, limiting insights into population-level representations, while, the role of correlation in neural coding remains a subject of considerable debate. To address this, we introduce novel, correlation-aware information-theoretic measures that quantify the encoding efficiency of multiple neurons, including the joint stimulus information rate for neuron pairs and the spectral-stimulus information for arbitrary sized populations. The spectral-stimulus information, defined as the leading eigenvalue of the stimulus information matrix, is maximized when neurons exhibit localized, non-overlapping firing fields, mirroring place cell and head direction cell activity. We apply these measures to neural data recorded in mice and monkeys,  elucidating differences in encoding efficiency across neuronal pairs and populations. Then, we demonstrate that these measures can be used to train recurrent neural networks (RNNs) via self-supervised learning, leading to the emergence of place cells and head direction cells. Our findings highlight how neural populations collectively encode stimuli, offering a more comprehensive framework for understanding stimulus encoding and optimizing artificial navigation systems in novel environments. 
\end{abstract}



\end{frontmatter}
%%%%%%%%%%%%%%%%%%%%%%%%%%%%%%%%%%%%%%%%%%%%%%
%% Please use \tableofcontents for articles %%
%% with 50 pages and more                   %%
%%%%%%%%%%%%%%%%%%%%%%%%%%%%%%%%%%%%%%%%%%%%%%
%\tableofcontents

\section{Introduction}

\subsection{Overview}

The brain possesses several specialized neurons that fire in response to given stimuli, such as location, head direction, speed, noise, and distance from walls [\cite{kropff2015speed, lever2009boundary, o1971hippocampus, taube1990head, wang2005sustained}]. The process of neurons firing in response to stimuli is known as neural encoding.

To measure the efficiency of encoding, neuroscientists often turn to information theory [\cite{shannon1948mathematical}], the mathematical study of the transmission and processing of information. These techniques have been used to determine how much information neurons carry about stimuli [\cite{dayan2005theoretical, quian2009extracting, skaggs1992information, brenner2000synergy, schneidman2003synergy, amari2006correlation, latham2013role, timme2014synergy}]. In particular, \cite{quian2009extracting} showed that the capacity of a set of neurons to decode position relates to the amount of spatial information conveyed by the neurons. Paramount in the relationship between information theory and neuroscience is the formulation of the spatial information rate by Skaggs et al. [\cite{skaggs1992information}]. The spatial information rate, which depends on a neurons firing rate over space, has seen tremendous use in characterizing how well single neurons encode space in both experimental studies [\cite{alme2014place, aronov2017mapping, fenton2008unmasking, finkelstein2015three, frank2000trajectory, fu2017tau, gardner2022toroidal, hayman2011anisotropic, nieh2021geometry, ormond2022hippocampal}] and in theoretical studies [\cite{souza2018information, wang2019place}]. While Skaggs was focused on how well a single neuron can encode space, other studies have introduced information-theoretic measures for multiple neurons with respect to arbitrary stimuli, which differ in subtle but important ways. In particular, the stimulus-specific information [\cite{butts2003much, butts2006tuning}] has been applied to quantify auditory encoding [\cite{kayser2010millisecond}]; synergy (or interaction information) [\cite{brenner2000synergy, schneidman2003synergy, timme2014synergy}] has been used to assess neural coding; and more recently, partial information decomposition (PID) [\cite{williams2010nonnegative}] has been developed to separate information into several intuitive components. A recurring issue across this literature is how to construct or approximate the multivariate response distribution when neurons exhibit correlated activity: obtaining this joint distribution is computationally nontrivial and often requires assuming a parametric model [\cite{amari2006correlation, latham2013role}]. As a result, even though correlations may influence population coding, their role remains context-dependent and challenging to incorporate in practice [\cite{amari2006correlation, schneidman2003synergy, series2004tuning, latham2013role, butts2006tuning}]. 

In recent years, there has been significant interest in building computational models that mimic mammalian spatial representation, as a proxy for understanding the utility of underlying neural dynamics [\cite{banino2018vector, burak2009accurate, couey2013recurrent, cueva2018emergence, fuhs2006spin, guanella2007model, mcnaughton2006path, mitchell2024topological, nasrin2019bayesian, schaeffer2024self, sorscher2019unified, sorscher2023unified, sengupta2018manifold, pettersen2024self, pettersen2024learning, issa2012universal, redman2024not, whittington2020tolman}]. Many implementations train recurrent neural networks (RNNs) for path integration via supervised learning. These models are trained to predict position or to match their output to that of predetermined place cells [\cite{cueva2018emergence, sorscher2019unified, sorscher2023unified}]. Self-supervised learning, which does not rely on external targets, has just recently been applied to the brain’s spatial representations and is only moderately well explored in neuroscience literature [\cite{zbontar2021barlow, schaeffer2024self, raju2024space}].

Here, we introduce novel information-theoretic measures to quantify the efficiency of neural population encoding and use them to train RNNs to encode stimuli through self-supervised learning. The proposed measures include the joint stimulus information rate for two neurons and the spectral-stimulus information for an arbitrary number of neurons. These extend classical rate-based information metrics by introducing correlation explicitly into the analytic structure of the joint information rate without assuming activity or conditional independence. In this way, the formulation bridges firing correlations and stimulus coding in a closed-form, data-driven expression that yields a correlation-aware measure of coding efficiency and naturally extends to arbitrary population sizes. The spectral-stimulus information is defined as the leading eigenvalue of the stimulus information matrix, whose entries are joint stimulus information rates. We prove that that this measure is maximized when neurons exhibit localized, minimally overlapping tuning, akin to place cell activations across space and head direction cell tuning across angles. Together, these metrics capture both stimulus selectivity and anti-correlation in population codes, yielding a compact, interpretable measure of encoding efficiency. Finally, we show that our metrics can be used to train RNNs to encode stimuli through self-supervised learning. In particular, our models are capable of learning many place cells and head direction cells.

\subsection{Place and Head Direction Cells}
\begin{figure}
    \centering
    \includegraphics[width = 0.8\linewidth]{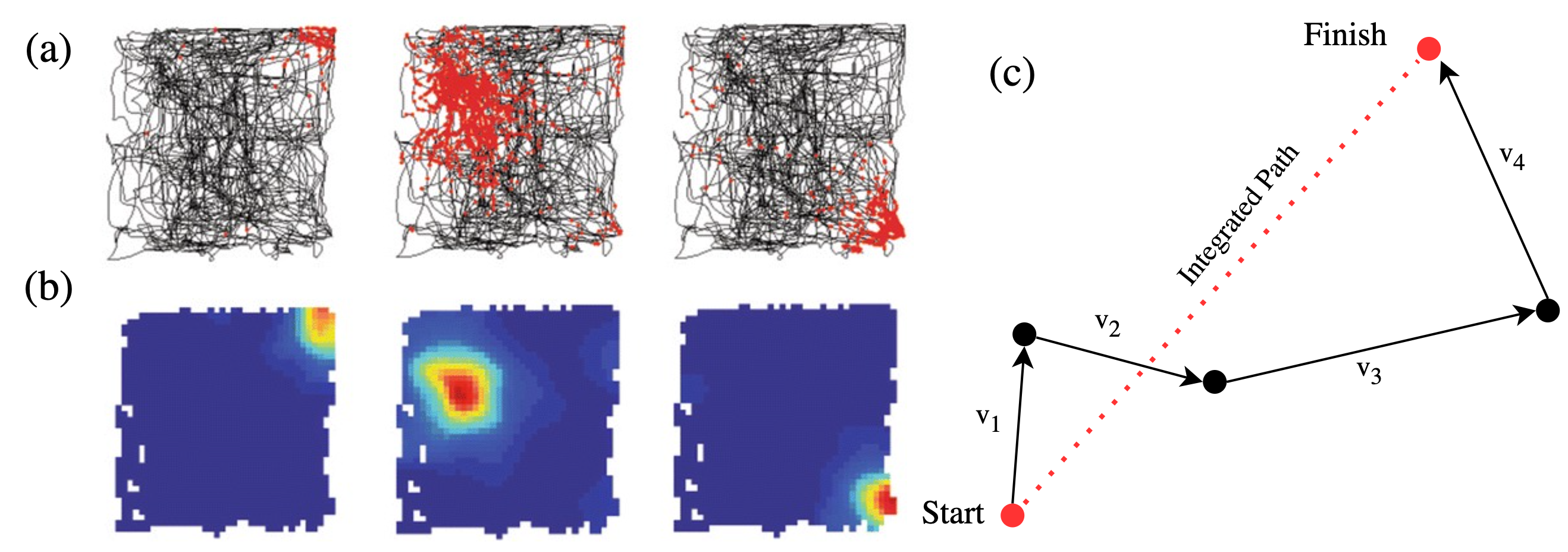}
    \caption{(a) Place cells recorded in hippocampal subarea CA3. Firing locations are shown as red dots on the path of a rat (black). (b) Activity maps of the cells. Red is high firing rate, and blue is no firing. (c) Diagram of the process of path integration. The agent integrates four velocities to determine its current location via the integrated path (red). (a-b) adapted from \cite{stensola2016grid}.}
    \label{place_cell_path_integration}
\end{figure}

\cite{o1971hippocampus} discovered spatially selective neurons in the hippocampus, which they termed place cells. Place cells form a sparse representation of space by firing at one or several seemingly random locations in small and large environments [\cite{rich2014large, harland2021dorsal}] (Figure \ref{place_cell_path_integration} (a) and (b)). The firing fields of place cells are known to create a sparse representation of space, with consistent activity responses in the absence of additional stimuli [\cite{save1998spatial}], a property we refer to as uniformity. \cite{hafting2005microstructure} discovered another set of spatially selective neurons in the medial entorhinal cortex (MEC), termed grid cells. In contrast to place cells, grid cells exhibit regular hexagonal spatial firing fields. Together, the MEC and hippocampus can encode relative spatial location, without reference to external cues, by integrating linear and angular self-motion [\cite{mcnaughton2006path}]. This process is described as path integration or dead reckoning (Figure \ref{place_cell_path_integration} (c)).  

Studies in mice and rats have shown that place cells use several factors when determining where to fire in space, including self-motion and the presence of boundaries [\cite{chen2013vision, lever2002long, o1976place}]. In 2017, Keinath et al. [\cite{keinath2017environmental}] showed that the hippocampal map realigns after disorientation, preserving place cell anti-correlated firing, even after realignment within the environment. This suggests that when seeking to understand how place cells encode space, analyzing a pair or population of place cells rather than each one individually is more useful, as they collectively act to form a cognitive map [\cite{o1978hippocampus, quian2009extracting}]. 

Head direction cells encode the direction the head is facing at any moment in time, independent of the animal’s position or body orientation [\cite{taube1990head}]. These neurons fire maximally when the animal’s head is oriented toward a specific “preferred direction,” providing a stable internal representation of heading that remains consistent across different behavioral contexts and even in darkness [\cite{yoder2015visual}]. Similar to place cells, head direction cells within a population exhibit structured relationships: neurons with opposing preferred directionalities show strongly anti-correlated firing patterns, reflecting the circular nature of directional encoding [\cite{peyrache2015hd, angelaki2020head}]. This coordinated activity indicates that the neural representation of heading emerges from interactions across the entire population rather than independent tuning of individual cells.

Information-theoretic measures that treat neurons as isolated encoders risk missing critical interactions that underlie stable spatial and directional representations. Due to the collective nature of neural activity, there is a clear and distinct need for rate-based metrics that can accurately evaluate the encoding of populations of neurons without assuming neural independence.

\subsection{Computational Models of Mammalian Navigation}

By training RNNs to perform path integration based on velocity inputs, grid cell-like representations may form in the hidden layer [\cite{banino2018vector, cueva2018emergence, schaeffer2024self, sorscher2019unified}]. This suggests that grid cells, border cells, and other cells observed in MEC, are a natural solution for representing space efficiently given the predominant recurrent connections in the neural circuits [\cite{cueva2018emergence}]. However, it has been recently suggested that the emergence of grid cells in RNNs with velocity inputs may be strongly driven by particular, non-fundamental, and post-hoc implementation choices as opposed to fundamental truths about neural circuits or the loss function(s) they optimize [\cite{schaeffer2022no}]. That is, grid cells are not the unique solution for space representation, but perhaps an unexpected solution to the path integration problem [\cite{issa2012universal}]. In fact, grid cells can even be learned in spatial encoding tasks without path integration [\cite{pettersen2024self}] and are less prominent in RNNs trained to path integrate more than one agent [\cite{redman2024not}]. Recent normative modeling approaches suggest grid cells construct a basis for predicting future outcomes [\cite{stachenfeld2017hippocampus, yu2020prediction}], and encode not only variables of interest but also how the variable transforms [\cite{dorrell2022actionable}].

In addition to grid cells, place cells have begun to receive attention from the computational neuroscience and machine learning communities [\cite{pettersen2024learning, whittington2020tolman, wang2024time, sun2025learning, raju2024space, mainali2025universal, eliav2021multiscale, malerba2025random}]. It has been demonstrated that place cell like responses can be learned via similarity-based objectives [\cite{sengupta2018manifold, pettersen2024learning}], and that irregular place cell responses in large environments can be reproduced via a random Gaussian process [\cite{kingma2014adam}]. Furthermore, the Tolmen-Eichenbaum Machine, which is an unsupervised model trained to predict the next sensory experience from all previous sensory experiences, is capable of learning grid, place, and border cell-like representations [\cite{whittington2020tolman}].

\subsection{Outline and Summary of Findings}

The remainder of this paper is organized as follows. Section \ref{sec:stim_info_rates} provides a brief review of stimulus information rates and introduces our novel metrics, which used to elucidate responses of both pairs and populations of biological neurons (Section \ref{subsec:real_dat}). Section \ref{sec:SSSE} details the process in which RNNs are trained to encode stimuli via self-supervised learning with spectral-stimulus information or correlation-agnostic models; the results of these experiments are presented in Section \ref{place cell score results}. Section \ref{sec:results} contains the main results and additional analyses of learned activations. Finally, we briefly discuss our findings and propose future directions in Section \ref{sec:discussion}. The main contributions of this work are as follows:
\begin{itemize}
    \item We prove that the joint stimulus information rate and spectral–stimulus information are maximized when neurons exhibit high specificity and anti-correlation, suggesting that information is most efficiently encoded when neurons fire in distinct, localized regions of the stimulus domain.
    \item We show that our metrics capture meaningful changes in place cell encoding across different environments. 
    \item We find that RNNs trained with spectral–spatial information develop a greater number of high-quality place cells than networks trained using Skaggs' spatial information rate, allowing for more accurate decoding.
\end{itemize}

\section{Stimulus Information Rates}
\label{sec:stim_info_rates}
In this work, we will be interested in information regarding some stimulus space $S$ representing the set of possible values a stimulus can take. For example, $S$ may be locations, directions, or speeds. Focusing on discrete stimulus spaces, the \textit{occupancy distribution} is the probability mass function such that $p(s)$ for $s \in S$ is the probability that stimulus $s$ occurs.

Stimulus information rates [\cite{brenner2000synergy}] depend only on the occupancy distribution and the neuron firing rates across the stimulus space. To obtain the stimulus information rate of a neuron $A$ for stimulus space $S$, we model the probability of neuron $A$ firing during a short time interval $\Delta t$ as a Bernoulli distribution with parameter $\lambda_A(s)\Delta t$ where $\lambda_A(s)$ is the mean firing rate of neuron $A$ given stimulus $s \in S$.  
The \textit{stimulus information rate} [\cite{brenner2000synergy, skaggs1992information}] (in bits/s) of a single neuron $A$, across stimulus space $S$, is
\begin{align*}
    I_{sec}(A: S) = \sum_{s \in S}\lambda_A(s) \log_2 \left(\frac{\lambda_A(s)}{\bar{\lambda}_A}\right) p(s),
\end{align*}
where $p(s)$ is the probability of the agent receiving stimulus $s,\ \lambda_A(s)$ is the mean firing rate of the neuron $A$ with stimulus $s$, and $\bar{\lambda}_A = \sum_{s \in S}\lambda_A(s) p(s)$ is the mean firing rate of neuron $A$ over the entire stimulus space [\cite{brenner2000synergy}]. Table \ref{table:mathematical notation} in the Supplementary Materials lists the common mathematical notation used throughout this work. 

Measuring the information quantity in bits/spike is often more useful for distinguishing information content. This is denoted as $I_{spike}$, and defined by 
\begin{align}
\label{spatial information rate}
    I_{spike}(A: S) = \frac{1}{\bar{\lambda}_A} I_{sec}(A:S)= \sum_{s \in S} \frac{\lambda_A(s)}{\bar{\lambda}_A} \log_2 \left(\frac{\lambda_A(s)}{\bar{\lambda}_A}\right) p(s),
\end{align}
$I_{spike}$ measures the \textit{specificity} of a neuron with respect to stimulus $S$ [\cite{skaggs1992information}]. 

\begin{figure}
    \centering
    \includegraphics[width=0.8\linewidth]{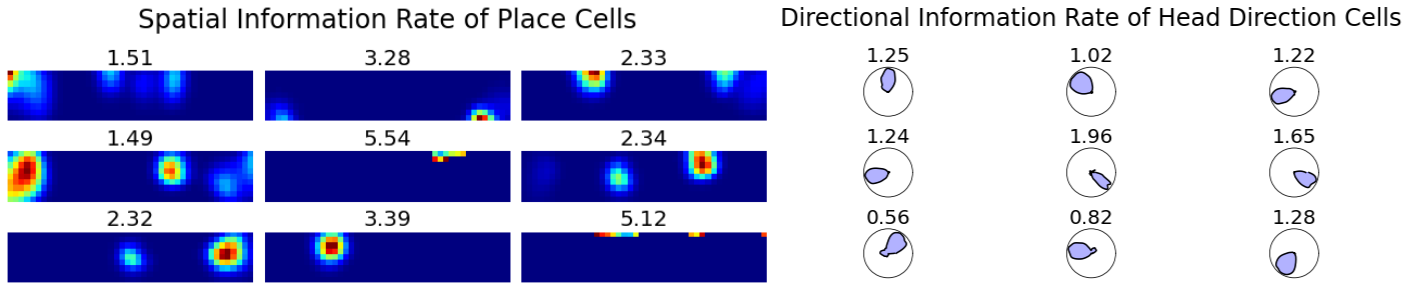}
    \caption{Example of the spatial information rate (bits/spike) of place cells recorded in [\cite{hazama2019data}] and directional information rates (bits/spike) of head direction cells recorded in [\cite{peyrache2015hd}]. Both scenarios use Equation \ref{spatial information rate}, albeit with different stimuli $S$.}
    \label{fig: bioinfos}
\end{figure}

As demonstrated in Figure \ref{fig: bioinfos}, if $S$ represents a set of spatial locations, we refer to the stimulus information rate as the spatial information rate. Similarly, if $S$ represents a set of orientations, we refer to it is as the directional information rate. This nomenclature applies to all future information measures in this work, in which ``stimulus" may be replaced with the specific variable whose encoding by a neuron we wish to evaluate.

Theorem \ref{thm:max_stiminfo} proposes that the stimulus information rate is maximized when a neuron fires only in response to a single stimulus $s \in S$. The proof of Theorem \ref{thm:max_stiminfo} and those of the rest of the new theoretical considerations are delegated in Supplementary Materials  \ref{app:proofs}.

\begin{theorem} For a fixed occupancy distribution such that $\min_i p_i = \epsilon$, the maximum of the stimulus information rate (in bits/spike) is $\log_2(1/\epsilon)$. For a uniform occupancy distribution with $n_s$ stimulus bins, the maximum of the stimulus information rate (in bits/spike) is $\log_2(n_s)$. 
\label{thm:max_stiminfo}
\end{theorem}

\begin{figure}
    \centering\includegraphics[width=0.8\linewidth]{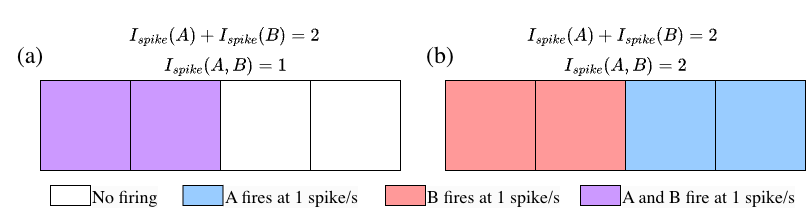}
    \caption{Example of two neurons with correlated (a) and anti-correlated (b) firing patterns. While the joint stimulus information rate can distinguish between the two scenarios, naive combinations of the individual stimulus information rates yield the same result.}
    \label{fig:jointvsskaggs}
\end{figure}

Although the stimulus information rate effectively gauges a single neuron's encoding efficiency, it is poorly suited for analyzing pairs or populations of neurons. For example, the sum of the stimulus information rates of two neurons is the same regardless of whether or not they have correlated or anti-correlated firing patterns (Figure \ref{fig:jointvsskaggs}). To accommodate this, we define the joint stimulus information rate of two neurons. 

\begin{definition}
\label{JSI def}
Consider two neurons $A$ and $B$, and let $r \in [-1, 1]$ be the Pearson's correlation coefficient between neurons over some stimulus domain $S$. Without assuming activity or conditional independence, we can explicitly write the joint distribution for neurons $A$ and $B$ firing (or not firing) over a short time frame $\Delta t$. Then, if $p$ is an occupancy distribution over $S$, then the \textit{joint stimulus information rate} (in bits/s) of neurons $A$ and $B$ is
\begin{align*}
    I_{sec}(A,B: S) &= \sum_{s \in S} \Bigg[ p(s) r\sqrt{\lambda_{A,B}(s)}\log_2\left(\frac{\sqrt{\lambda_{A,B}(s)}}{\tilde{\lambda}_{A,B}} \right) \nonumber \\ 
    &+ p(s)\left(\lambda_A(s) - r\sqrt{\lambda_{A,B}(s)}\right)\log_2\left(\frac{\lambda_A(s) - r\sqrt{\lambda_{A,B}(s)}}{\bar{\lambda}_A - r\tilde{\lambda}_{A,B}}\right)  \\
    &+ p(s)\left(\lambda_B(s) - r\sqrt{\lambda_{A,B}(s)}\right)\log_2\left(\frac{\lambda_B(s) - r\sqrt{\lambda_{A,B}(s)}}{\bar{\lambda}_B - r\tilde{\lambda}_{A,B}}\right) \Bigg], \nonumber 
\end{align*}
where $\lambda_A(s)$ and $\lambda_B(s)$ are the mean firing rates
of the neuron $A$ and $B$, respectively with stimulus $s$, $\lambda_A(s)\lambda_B(s) = \lambda_{A,B}(s)$, and $\sum_{s\in S} p(s) \sqrt{\lambda_{A,B}(s)} = \tilde{\lambda}_{A,B}$. 

We can likewise define $I_{spike}(A, B:S)$ by choosing an appropriate normalization term:
\begin{align*}
%\label{I_spike(A,B)}
    I_{spike}(A,B: S) = \frac{1}{\Lambda_{A, B}} I_{sec}(A,B),
\end{align*}
where $\Lambda_{A,B} = \sum_{s \in S} p(s) \frac{\lambda_A(s) + \lambda_B(s)}{2} = \frac{\bar{\lambda}_A + \bar{\lambda}_B}{2}$.
As desired, $I_{spike}(A, B:S)$ is in units of bits/spike. Furthermore, as shown in Proposition \ref{properties} of Supplementary Materials \ref{app:proofs}, $I_{spike}(A, B:S) = I_{spike}(B, A:S)$, and $I_{spike}(A, A:S) = I_{spike}(A:S)$. The full derivation of the joint stimulus information rate is included in Construction \ref{const JSIR}. 

\end{definition}
When two neurons do not fire in the same part of the stimulus domain, the joint stimulus information rate is proportional to the sum of the individual stimulus information rates of the neurons, as described in Proposition \ref{prop:no-overlap}.

The joint stimulus information rate can quantify the efficiency of a pair of neurons encoding a stimulus (Figure \ref{fig:jointvsskaggs}). That is, it prefers when neurons exhibit anti-correlated firing over the stimulus space.  Furthermore, it retains the desirable properties of the stimulus information rate—it is maximized when neurons fire in small regions of the stimulus space (Theorem \ref{thm:max_jointstiminfo}). 

\subsection{Redundancy-Synergy Index}

Equipped with the joint stimulus information rate, we use the principles of \cite{williams2010nonnegative} to decompose the information into several intuitive components, including redundancy (Rdn), unique information (Unq), synergy (Syn), and the redundancy-synergy index (RS). Using the joint stimulus information rate, these measures can be easily computed without assumptions about firing independence using only the firing rates of the neurons. For a complete derivation of all components, see Section \ref{PID_supplemetnary}. In this work, we will be mainly focused in the redundancy-synergy (RS) index, which we now define as the difference between the joint stimulus information rate and the sum of the individual stimulus information rates: 
\begin{align}
\label{eq:RS}
    \text{RS}(A,B:S) = I_{spike}(A,B:S) - I_{spike}(A:S) - I_{spike}(B:S).
\end{align}
In previous versions of RS, higher RS has been speculated to imply stimulus-dependent correlations, while negative values indicate redundancy [\cite{chechik2001group, brenner2000synergy, schneidman2003synergy}]. These interpretations hold in our case as well. In particular, if $A$ and $B$ have identical firing patterns (i.e., $A = B$), then $\text{RS}(A,B:S) = - I_{spike}(A:S)  = - I_{spike}(B:S)$. On the contrary, if $A$ and $B$ have anti-correlated firing patterns, then $\text{RS}(A,B:S) \approx 0$.

\subsection{Spectral Analysis}

The system of equations resulting from stimulus information rate techniques becomes underdetermined for more than two neurons. To overcome this and enable analysis of encoding efficiency across arbitrary population sizes, we construct and study matrices whose entries are pairwise joint stimulus information rates.

\begin{figure}
    \centering
    \includegraphics[width=0.8\linewidth]{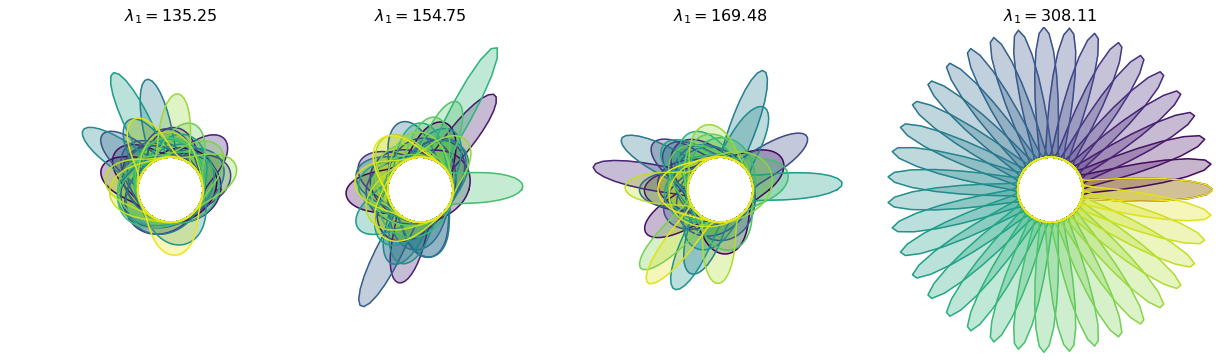}
    \caption{Leading eigenvalue of stimulus information matrix generated from four populations of 40 simulated Gaussian head direction cell firing patterns. The final scenario is evenly distributed firing fields. A larger leading eigenvalue indicates superior direction encoding.}
    \label{fig:lambda_1_hd}
\end{figure}

\begin{definition}
\label{stimulus information matrix}
The \textit{stimulus information matrix} $J \in \R^{n_p \times n_p}$ for $n_p$ neurons is the matrix whose entries are given by
\begin{align*}
%\label{equation for stimulus information matrix}
    J_{i, j} = I_{spike}(A_i, A_j: S).
\end{align*}
$J$ is symmetric, and its diagonal consists of the stimulus information rates of the individual neurons, $I_{spike}(A_i: S)$ for $i = 1,\ldots, n_p$.  If $\lambda_1,  \ldots \lambda_{n_p}$ are the eigenvalues of $J$ such that $|\lambda_1| \ge |\lambda_2| \ge,  \ldots, \ge|\lambda_{n_p}|$, we have that $\lambda_i \in \R$ for all $i = 1,\ldots, n_p$. Furthermore, if $J_{i,j}= I_{spike}(A_i, A_j: S) \ge 0$ for all $i,j$, then $\lambda_1 \ge 0$ by the Perron-Frobenius Theorem [\cite{gantmakher2000theory}]. We call this eigenvalue the \textit{spectral-stimulus information}. The spectral-stimulus information increases as neuron firing fields orthogonalize over the stimulus domain (Figure \ref{fig:lambda_1_hd}). As before, we may replace stimulus with the appropriate stimulus, such as spectral-spatial, or spectral-directional information, to reference the variable whose encoding by neurons we wish to evaluate. 
\end{definition}

Given a population of neurons, if no two neurons fire within the same region of the stimulus domain, the stimulus information matrix and spectral-stimulus information can be found explicitly in terms of the individual stimulus information rates (Proposition \ref{prop:no_intersect}). Proposition \ref{prop:no_intersect} extends Proposition \ref{prop:no-overlap} to the multi-neuron case. In the absence of overlapping firing fields, the stimulus information matrix (and corresponding spectral-stimulus information) depend solely on the stimulus information rates of the individual neurons. When the number of neurons matches the number of stimulus bins and no neuron is completely silent, we can provide an even more precise description. This special case is formalized in Theorem \ref{thm:special_case} and \ref{cor:eigenvalue_eigenvector}. Under these circumstances, the spectral-stimulus information is precisely $\lambda_1 = \log_2(n_s)(2n_p - 1)$ and its corresponding eigenvector is $\textbf{1} \in \R^{n_p}$, the vector of all ones.

One should expect the spectral-stimulus information to be maximized when all neurons fire in non-overlapping regions of the stimulus domain. That is, given a stimulus domain with $n_s$ stimulus bins and $n_p = n_s$ neurons, it is best that each neuron fire in its unique stimulus bin. Indeed, this is true, and we prove this result in Theorem \ref{thm:largest_eigenvalue}.

\begin{theorem} Given a uniform occupancy distribution and $n_p = n_s$ neurons, the spectral-stimulus information is maximized when $\{s\in S: \lambda_{A_i}(s) > 0\} \neq \emptyset$ for all $i = 1,\ldots, n_p$ and $\{s\in S: \lambda_{A_i}(s) >0\} \cap \{s\in S: \lambda_{A_j}(s) > 0\} = \emptyset$ for all $i, j = 1,\ldots, n_p,\ i\neq j$.
    \label{thm:largest_eigenvalue}
\end{theorem}

Theorem \ref{thm:largest_eigenvalue}, together with empirical evidence (Figure \ref{fig:lambda_1_hd}, Figure \ref{fig:diss_evect}), motivate the use of spectral-stimulus information for self-supervised encoding, which we describe in the following section. It suggests that training neurons to optimize spectral-stimulus information can result in anti-correlated firing across pairs of neurons, similar to that of place and head direction cells. 

\section{Self-Supervised Stimulus Encoding}
\label{sec:SSSE}

Self-supervised stimulus encoding refers to training neurons to represent a particular feature of the sensory environment without external targets. Here, we encode a stimulus into RNN output neurons via information theory. 

\subsection{Recurrent Neural Networks for Stimulus Encoding}

\begin{figure}
    \centering
    \includegraphics[width = 0.8\linewidth]{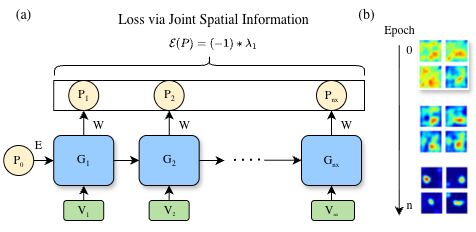}
    \caption{(a) Diagram of RNN architecture used to perform self-supervised learning. Shown here across $n_x$ steps. The internal state is updated via Equation \eqref{eq:hidden_state} and outputs are produced according to Equation \eqref{eq:output_state}. (b) As training progresses over $n$ epochs, the neuron's firing fields become more place cell-like. Here the firing fields of four neurons are shown.}
    \label{fig:rnn_architecture}
\end{figure}

Recurrent networks are a class of artificial neural networks that process data sequentially. These networks maintain an internal representation, known as the hidden state, which is recursively updated as the network receives inputs. For an Elman RNN [\cite{elman1991distributed}] with a sequence of input data $\{v_1, \ldots, v_{n_x}\}$, the internal state is updated via
\begin{align}
\label{eq:hidden_state}
    G_t = \sigma(W_{ih}v_t + W_{hh}G_{t-1}),
\end{align}
where $G_t$ is the hidden state vector, $W_{ih}$ and $W_{hh}$ are the input and hidden layer weight matrices, respectively, and $\sigma$ is the activation function, such as $ReLU(\cdot)$ (Figure \ref{fig:rnn_architecture}). The output of the network at time $t$ is given by 
\begin{align}
\label{eq:output_state}
    P_t = \sigma(WG_t).
\end{align}
For self-supervised stimulus encoding, our data comes as alterations to the stimulus space. For location in two dimensions, the inputs are velocities in two dimensions. For head direction, input data takes the form of one-dimensional angular velocities. The initial activations $P_0 \in \mathbb{R}^{n_p}$ are chosen by calculating distances from uniformly randomly sampled locations in the arena and applying a difference of softmaxes. This process is explained in detail in the Supplementary Materials \ref{experimental_details}. The initial activations are then encoded by $E$ into the hidden state. 

We use the stimulus information matrix $J$ (Definition \ref{stimulus information matrix}), associated with neuron output activations $P = [P_1, \cdots, P_{n_x}]^T$ to analyze global information and calculate epoch loss (Figure \ref{fig:rnn_architecture} (a)). In detail, each column of $P$ serves as a distribution of firing rates $\{\lambda(s)\}_{s\in S}$ across the trajectory, from which spectral-stimulus information can be calculated. To learn place or head direction cells via $J$, one could choose to optimize the entries of $J$ or spectral-stimulus information, $\lambda_1$. These values are highly correlated (Remark \ref{evals}), and optimizing along either choice leads to similar outcomes (Figure \ref{fig:summation_vs_eigenvalue}). For our experiments, we choose to focus on the optimization of the spectral-stimulus information given its potential for geometric interpretation. That is, to increase information efficiency, we minimize 
\begin{gather}
\label{HO Loss}
    \mathcal{E}_{spectral}(P) = (-1)*\lambda_1,    
\end{gather}
where $\lambda_1$ is the leading eigenvalue of $J$. The joint stimulus information rate, $I_{spike}$, depends on the firing rates of the neurons, $P$, and the occupancy distribution, which is the probability mass function across $n_x$ steps along the trajectory. For simplicity, we assume this to be a discrete uniform distribution for all trajectories in the stimulus domain.

As a baseline, we compare the results of maximizing spectral information (Equation \eqref{HO Loss}) with the results of maximizing single-order information according to Skaggs' stimulus information rate (Equation \ref{spatial information rate}). This latter loss function is given by, 
\begin{gather}
\label{Skaggs Loss}
    \mathcal{E}_{skaggs}(P) = (-1)*\sum_{i = 1}^{n_p}\ I_{spike}(P[:,i]),
\end{gather} where $P[:,i] \in \R^{n_x}$ is the $i$th column of $P$. 
Additional experimental details, including network hyperparameters, are listed in Supplementary Materials \ref{experimental_details}.

\section{Results}
\label{sec:results}
\subsection{Spectral–Spatial Information and Eigenstructure of Population Coding Across Environments}
\label{subsec:real_dat}

\begin{figure}
    \centering
    \includegraphics[width=0.9\linewidth, keepaspectratio]{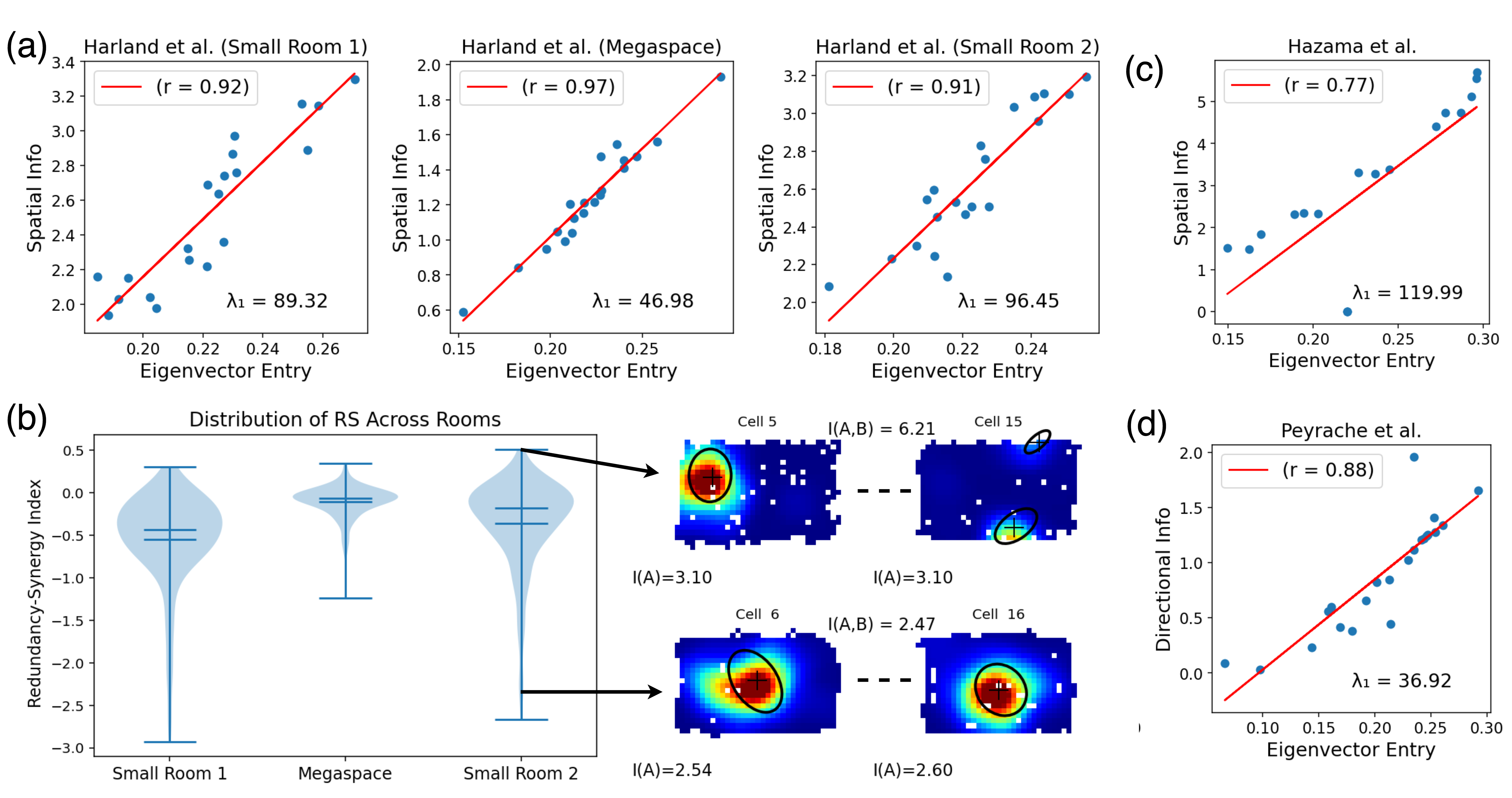}
    \caption{(a) Spatial information rate versus eigenvector entry magnitude from place cell recordings done in [\cite{harland2021dorsal}], including spectral-spatial information. (b) Distribution of Redundancy-Synergy Index (Equation \ref{eq:RS}) from place cell pairs recorded in [\cite{harland2021dorsal}] with two example pairs of place cells. Spatial information rate of each cell is reported, as is the joint spatial information rate for each pair of cells. (c) Same as (a), except with place cell recordings done in [\cite{hazama2019data}]. (d) Directional information rate versus eigenvector entry magnitude from head cell recordings done in [\cite{peyrache2015hd}], including spectral-directional information.}
    \label{fig:diss_evect}
\end{figure}

We analyzed a population of 20 place cells recorded in a small room ($< 4 \ m^2$) and a megaspace ($18.6\ m^2$) recorded in [\cite{harland2021dorsal}]. The spectral-spatial information was $89.32$ using data in the first small room recording, $46.98$ using data from the megaspace, and $96.45$ using data in the second small room recording (Figure \ref{fig:diss_evect} (a)). In the megaspace, the majority of place cells expressed multiple place subfields (2-5) of different sizes, with fields covering 5-20\% of the environment. In the small rooms, place cells typically had 1-2 place fields, covering 5-15\% of the environment. In addition, there was a greater degree in firing field overlap in the megaspace compared with the small room. Together, these properties lead to the discrepancy in spectral-spatial information between recordings from the small rooms and the megaspace, in which spectral-spatial information from small rooms is much larger compared to that of the megaspace.

We also computed the redundancy–synergy (Equation \ref{eq:RS}) for all pairs of place cells within each room (Figure \ref{fig:diss_evect} (b)). While most pairs of neurons exhibited near-zero redundancy–synergy, some pairs showed high redundancy. For instance, for two place cells with spatial information rates of 2.54 and 2.60 bits/spike, respectively, and overlapping firing fields, the joint spatial information rate was 2.47 bits/spike—comparable to the individual rates. Accordingly, the redundancy–synergy value for this pair was negative, reflecting substantial redundancy. In contrast, two place cells with non-overlapping firing fields, each with a spatial information rate of 3.10 bits/spike, exhibited a joint spatial information rate of 6.21 bits/spike—approximately the sum of their individual rates, indicating minimal redundancy and anti-correlated firing. These examples empirically illustrate how the joint spatial information rate is modulated by firing field overlap.

Corollary \ref{cor:eigenvalue_eigenvector} and Theorem \ref{thm:largest_eigenvalue} suggest that, under ideal conditions, stimulus information is uniformly distributed across neurons, corresponding to a leading eigenvector of the stimulus information matrix with equal entries. In empirical data, however, this uniformity seldom holds (Figure \ref{fig:diss_evect} (a), (c), and (d)). Instead, the leading eigenvector typically exhibits entries of varying magnitude, indicating unequal contributions of informations. For populations of place cells recorded in mice and monkeys, as well as head direction cells in mice, the stimulus information rate of each neuron is strongly positively correlated with its associated entry in the leading eigenvector of the stimulus information matrix. An illustrative example of this relationship is shown in Figure \ref{fig:summation_vs_eigenvalue} (b). Thus, the leading eigenvalue quantifies the overall efficiency of population-level encoding, while the corresponding eigenvector elucidates which neurons contribute most prominently to that encoding.

\subsection{Place cells arise via self-supervised learning}
\label{place cell score results}
RNNs receiving velocity inputs can produce place cell-like responses via self-supervised learning with information theory. These responses can be achieved by training the model to maximize either the eigenvalue of the spatial information matrix (Equation \ref{HO Loss}), which we refer to as spectral-spatial information, or the sum of the Skaggs' spatial information rates (Equation \ref{Skaggs Loss}). Here, we compare the techniques by analyzing the learned models and their corresponding encoding of space.

\begin{figure}
    \centering
    \includegraphics[height=0.33\textheight, keepaspectratio]{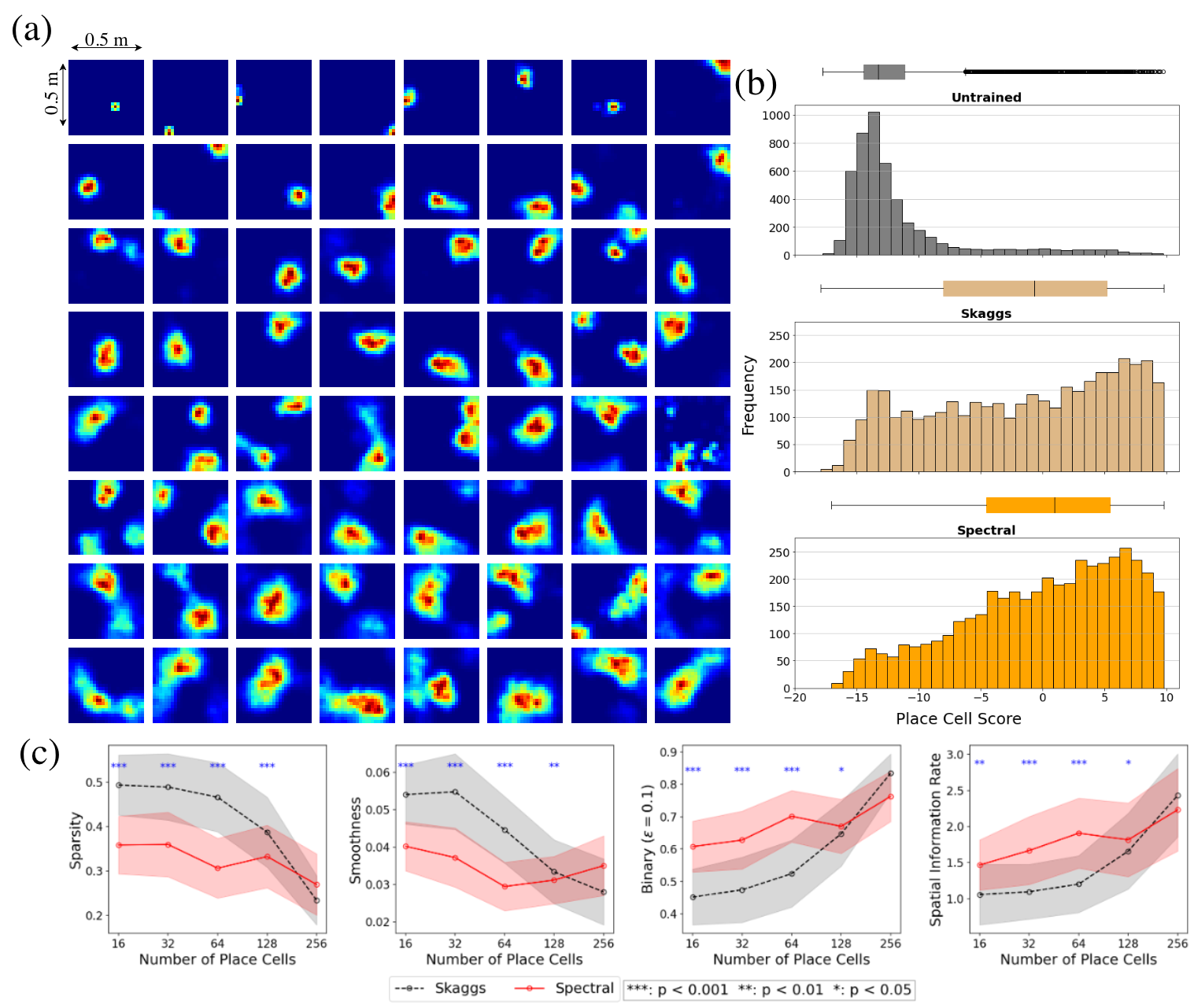}
    \caption{ (a) Activation maps of top 64 learned place cells from a model trained with spectral-spatial information, normalized to be between 0 and 1, and sorted by place cell score (Equation \eqref{place cell score} of \ref{app:pc_score}). (b) Original and trained place cell score box plots and histograms. Place cell scores are recorded across 50 total models for each loss function with varying place cell sizes. (c) Mean ($\pm$ standard error) sparsity, smoothness, binarity ($\epsilon = 0.1$), and spatial information rate (Equations \ref{smoothness}- \ref{arena spatial information} of \ref{app:pc_score}) of trained place cells as the number of place cells increases.} 
    \label{fig:top64}
\end{figure}

Training with spectral-spatial information significantly improves spatial encoding by enhancing the quantity and quality of place cells. Figure \ref{fig:top64} (a) shows 64 learned place activations from a model of 128 neurons. Indeed, the neurons learn to fire in small regions, which generally are non-overlapping. Figure \ref{fig:all_activations} compares this model's 128 place cell activation maps to the activation maps of the untrained model. To compare place cells produced by optimizing along our spectral-spatial information measure with control experiments, we construct a place cell score, which we describe in detail in Supplementary Materials \ref{app:pc_score}. Our place cell score is a combination of three characterizing features of place cells: (i) smoothness of the firing rate; (ii) binary firing rates across space; and (iii) sparse activations In Figure \ref{fig:top64} (b-c), we compare place cell scores of place cells learned via spectral-spatial information to those learned via Skaggs' spatial information. 

Overall, neurons trained with spectral-spatial information yielded a greater median place cell score (0.971) than neurons trained with Skaggs (-0.656), while both drastically improve upon untrained neurons (-13.28) (Figure \ref{fig:top64} (b)). In a sensitivity analysis (Supplementary Materials \ref{app:sensitivity}), we show that spectral-spatial information consistently outperforms Skaggs spatial information in learning place-cell-like representations regardless of changes in training parameters (batch size, number of hidden units, etc.). These improvements underscore the effectiveness of including spectral-spatial information when optimizing neural representations of space.  Place cell activation maps from all models are available in the supplementary materials. 

It is worth noting, however, that while spectral-spatial information yields, on average, better spatial information, sparsity, binarity, and smoothness for models with 16, 32, 64, and 128 neurons, there is no statistically significant difference for models trained with 256 neurons (Figure \ref{fig:top64} (c)). We speculated that this was the result of the size of the environment size, That is, 256 place cells may be too many to optimize within the $0.5\ m \times 0.5\ m$ environment; However, a second experiment revealed that these results persist even in larger environments (Figure \ref{fig:env_metrics_plot}). Despite these similarities, activations from 256-neuron models trained with spectral–spatial information produced stronger decoding performance than those trained with Skaggs’ spatial information (discussed in detail in Section \ref{subsec:decoding}).

\begin{figure}
    \centering
    \includegraphics[height = 0.33\textheight, keepaspectratio]{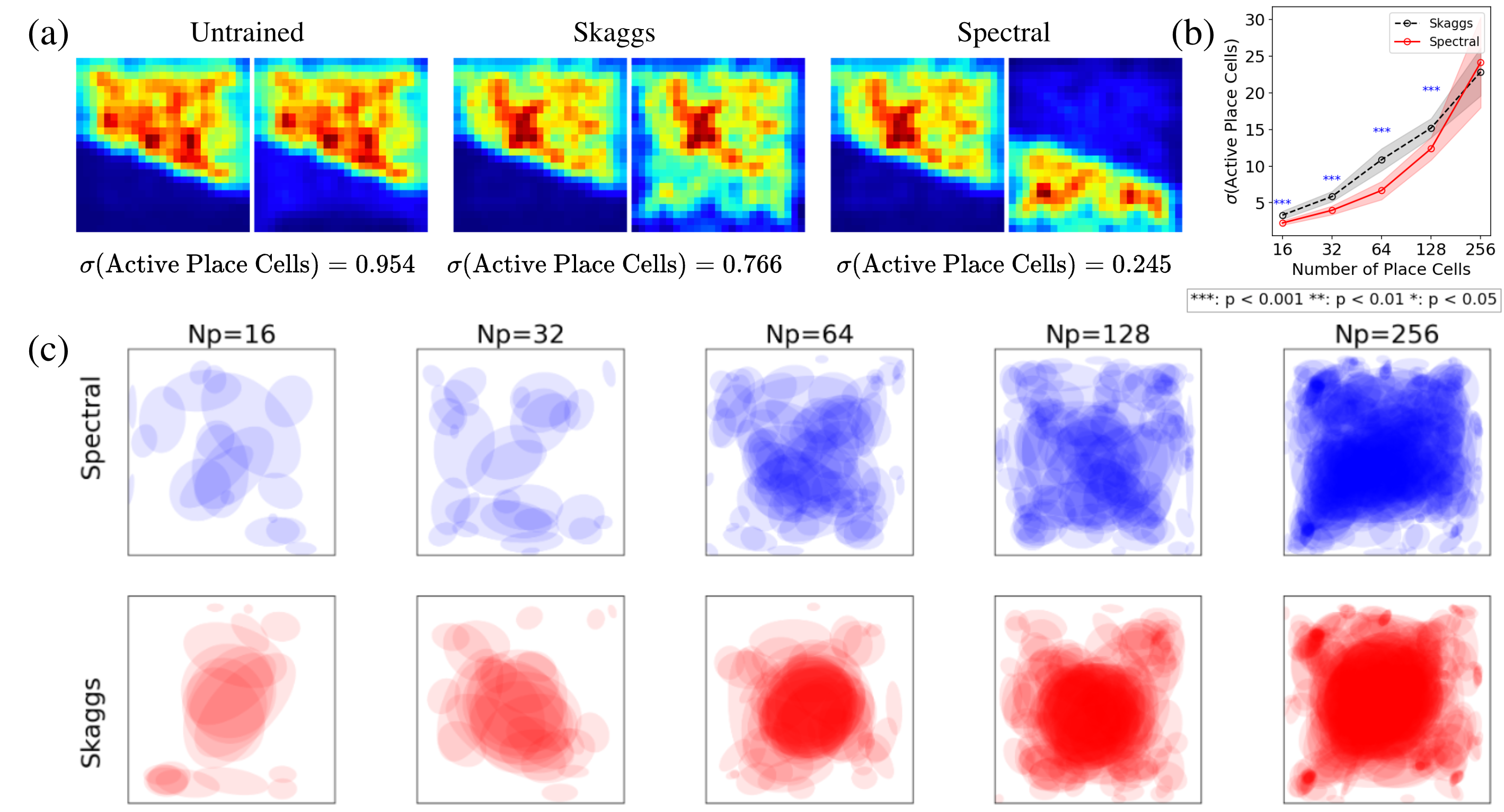}
    \caption{(a) Comparison of two place cell firing fields in which the initial cells fire in the same region. Standard deviation in active place cells decreases as firing fields separate. (b) Mean ($\pm$ standard deviation) of the standard deviation in the number of place cells with high activity (defined as exceeding 20\% of the peak firing rate of the rate map) as the number of trained place cells varies. (c) A comparison of the distribution of place fields between models trained with spectral-spatial information versus Skaggs.}
    \label{fig:simple_flip}
\end{figure}

Place cells learn anti-correlated firing only by training with spectral-spatial information (Equation \eqref{HO Loss}), as shown in Figure \ref{fig:simple_flip} (a). Training with the Skaggs loss function optimizes each cell's spatial information rate (Equation \eqref{Skaggs Loss}) and is agnostic toward the firing of the other cells. To measure the uniformity of learned spatial representations, we calculate the standard deviation in the number of highly active place cells at each discretized spatial region. A lower standard deviation suggests that the place cells have become more uniform as their firing fields disperse evenly across the domain.  

\begin{figure}
    \centering
    \includegraphics[width=0.9\linewidth]{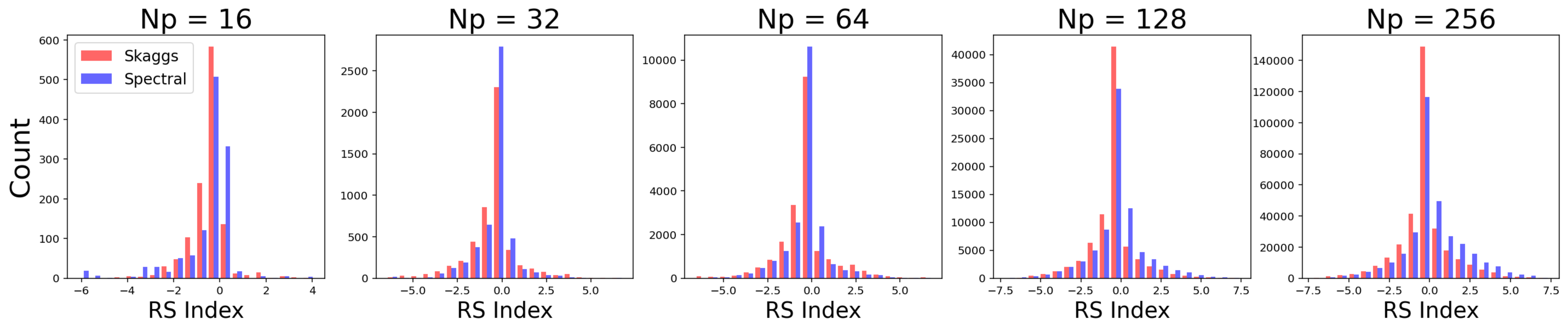}
    \caption{Redundancy-Synergy (RS) of pairs of  learned place cell activations from models trained with spectral-spatial information versus Skaggs' spatial information. Negative RS indicates redundancy. }
    \label{fig:rs}
\end{figure}

In Figure \ref{fig:simple_flip} (b), we show that place cells trained with spectral-spatial information result in greater uniformity than those trained with Skaggs' spatial information. For models trained with spectral-spatial information with 16, 32, 64, and 128 neurons, the standard deviation in the number of place cells with high activity across the spatial domain is lower. While, as with earlier measures, there is no statistically significant difference for models with 256 neurons. We further demonstrate this phenomena by comparing heatmaps of place cell fields from models trained with spectral-spatial information versus Skaggs' spatial information (Figure \ref{fig:simple_flip} (c)). In models trained with spectral-spatial information, place cell fields disperse throughout the environment, avoiding redundancy while maximizing information content. Models trained with Skaggs' spatial information result in many overlapping place fields and greater redundancy among pairs of neurons (Figure \ref{fig:rs}).

\subsection{Neural decoding}
\label{subsec:decoding}

\begin{figure}
    \centering
\includegraphics[width=\linewidth, keepaspectratio]{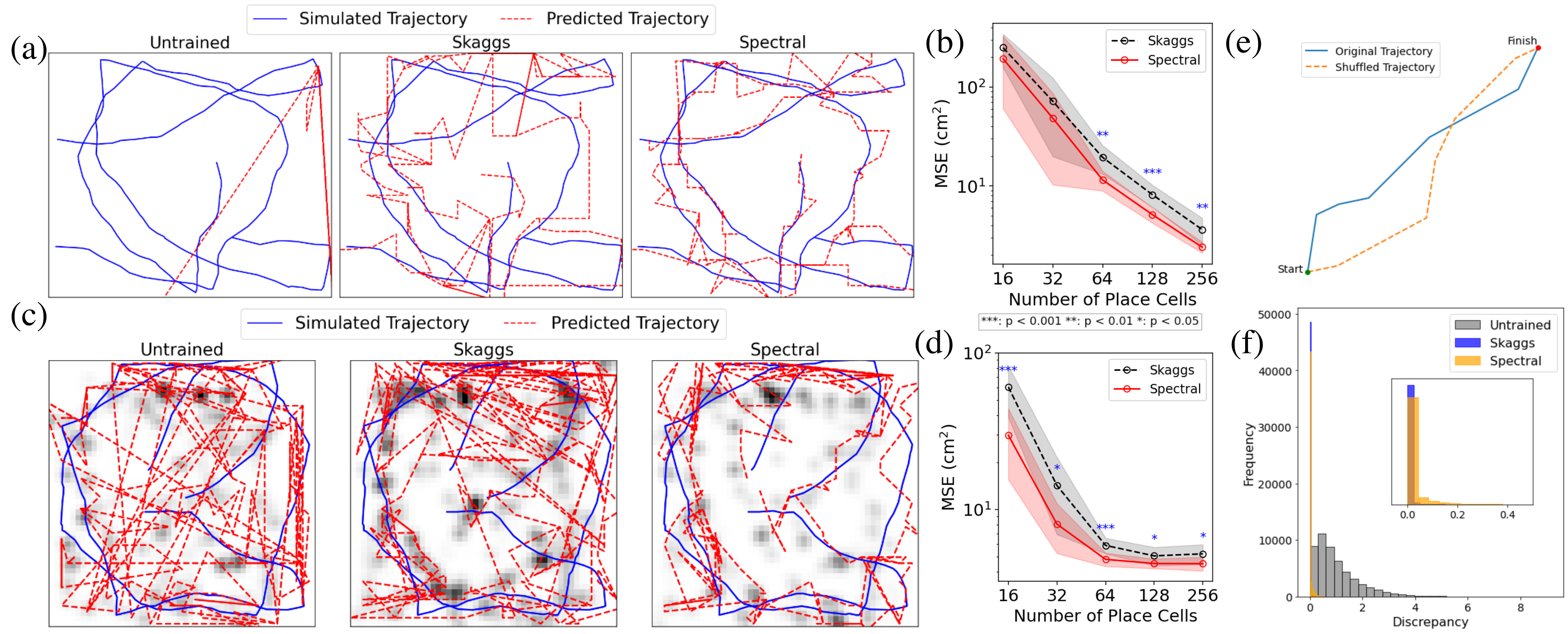}
    \caption{Comparison of decoding capabilities of neurons before and after training and path invariance. (a) Simulated and predicted trajectories leave-one-out classification with 128 neurons. (b) Skaggs vs spectral leave-one-out decoding mean square error (MSE) as the number of trained place cells increases. The mean ($\pm$ standard deviation) of MSE across experiments is shown. (c) Simulated and predicted trajectories Poisson-Bayes classification with 16 neurons, with posterior distribution included. (d) Skaggs vs spectral Poisson-Bayes decoding MSE as the number of trained place cells increases. The mean ($\pm$ standard deviation) of MSE across experiments is shown. (e) Two trajectories formed by shuffling the order of velocities. Since both trajectories end at the same location, neural representations at the final position should be identical despite different paths to get there. 
    (f) Histogram of path invariance discrepancy across 1000 trajectories from all untrained models, models trained with Skaggs' spatial information, and models trained with spectral-spatial information.}
    \label{fig:decoding}
\end{figure}

We consider two experiments to ensure the decoding abilities of the learned place cells. For each experiment, data from 10 trajectories (1000 positions) is used in the training or testing of the decoder. Each experiment is run once for each of the 50 models, i.e.\ 10 times for a given number of place cells, to obtain an average mean square error (MSE). We display the results in Figure \ref{fig:decoding}.

In the first experiment, we decode the agent's location from binarized place cell activations, one if the neuron firing rate is nonzero, and zero otherwise via leave-one-out classification. By turning the activations into binary values, spatial decoding must be done by relating a combination of active neurons to an area in the arena. Consequently, this decoding algorithm operates with the sole knowledge of which neurons are in an "active" state. In leave-one-out classification, one data point is reserved for testing and the remaining data points are used to train a Naive Bayes' Classifier (see [\cite{souza2018information}] for details). This procedure is repeated multiple times so that each firing rate value is used once for testing. 

Neurons trained with spectral-spatial information can predict the position with high precision, mirroring the simulated trajectory in all areas of the arena (Figure \ref{fig:decoding} (a)). Furthermore, these neurons consistently outperform neurons trained with Skaggs' spatial information rate (Figure \ref{fig:decoding} (b)). 

In the second experiment, we decode the agent’s position from the full place cell firing rates using a Poisson Bayesian decoder. To do so, we use the learned activations to simulate Poisson spikes from these rates, using the first 80\% of the data to estimate spatial tuning curves. For the remaining data, the decoder computes the posterior probability of the agent being in each spatial bin by combining the Poisson likelihoods of the observed spike counts across neurons. The decoded position is then taken as the maximum a posteriori (MAP) estimate, corresponding to the bin with the highest posterior probability. Averaging across the 50 trained models, neurons trained with spectral-spatial information produce lower mean squared error than neurons trained using Skaggs’ information rate (Figure \ref{fig:decoding} (c-d)). In particular, for 16 neurons, spectral-spatial training produced over a  50\% reduction in MSE relative to Skaggs’ neurons (29.9 cm$^2$ versus 60.6 cm$^2$).

\mycomment{
To generate Figure \ref{fig:decoding} (c-d), we train a support vector machine (SVM) with a linear kernel to predict which quadrant of the region the agent is in from the binarized neural input. Spectral-spatial information outperforms Skaggs spatial information by nearly 10\% when training just 16 neurons, and performs comparably for greater than 16 neurons. }

These experiments demonstrate that the place cells learned via self-supervised learning with spectral-spatial information are capable of better spatial decoding than neurons trained with Skaggs' spatial information. Combined with the results from Section \ref{place cell score results}, we conclude that self-supervised learning with spectral-spatial information allows for better spatial encoding with more place cell-like neural representations than previously attainable via Skaggs' spatial information. 

\subsection{Path invariance}

A key feature of any collection of neurons encoding spatial representation is its path invariance [\cite{schaeffer2024self}]. That is, two trajectories that begin and end at the same position should have identical final neural representations (Figure \ref{fig:decoding} (e)). To test for path invariance of our place cell representations, we form two trajectories by randomly shuffling a sequence of velocity vectors. % of the initial trajectory. T
We then compare the place cell activity by measuring the Euclidean distance of the discrepancy (Equation \eqref{path invariance discrepency}) between the two trajectories:  if $P_{n_x}, {P_{\text{shuffled}}}_{n_x} \in \R^{Np}$ are re-scaled activations at the final position, then the path invariance discrepancy $D$ is 
\begin{gather}
\label{path invariance discrepency}
    D = ||P_{n_x} -{P_{\text{shuffled}}}_{n_x}||_2^2.
\end{gather}
We measure $D$ for 1000 pairs of trajectories for each of the 50 models, and display results in Figure \ref{fig:decoding} (f). 

Our models trained with spatial information demonstrate exceptional performance (Figure \ref{fig:decoding} (f)). In detail, the models achieved an average ($\pm$ standard deviation) path invariance discrepancy of 0.004 $\pm$ 0.024 for Skaggs' optimization and 0.022 $\pm$ 0.066 for spectral optimization, while untrained models had a path invariance discrepancy of 1.024 $\pm$ 0.859. We view path invariance as a binary property that is present in both models trained with Skaggs' and spectral optimization (since their average path invariance discrepancy is small) and not present in untrained models. Altogether, path invariance further indicates that self-supervised learning combined with spatial information optimization results in models that successfully encode space. 

\subsection{Self-Supervised Head Direction Cell Learning}
\begin{figure}
    \centering
    \includegraphics[width=0.9\linewidth, keepaspectratio]{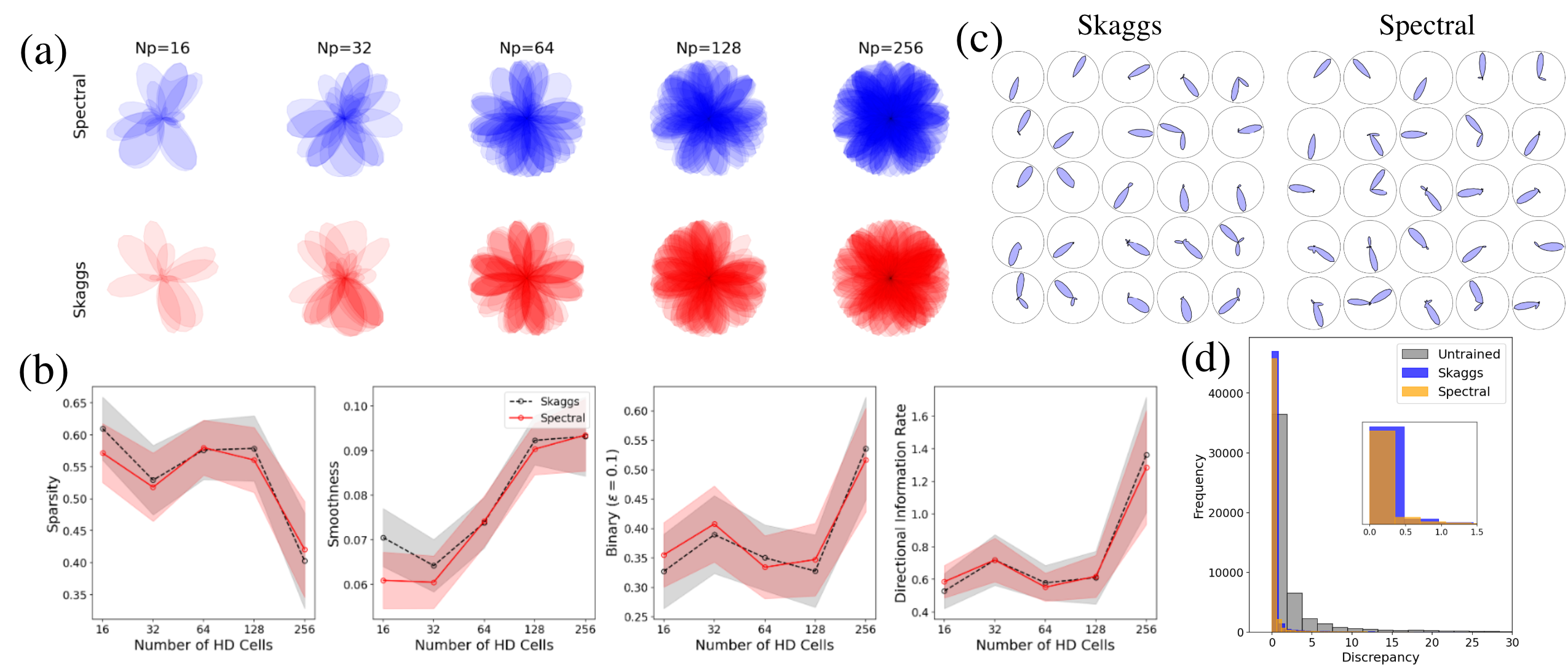}
    \caption{(a) Distribution of head direction cell firing fields of models trained with spectral-directional information versus Skaggs. (b). Mean ($\pm$ standard error) sparsity, smoothness, binarity ($\epsilon = 0.1$), and directional information rate (Equations \ref{smoothness}- \ref{arena spatial information} of \ref{app:pc_score}) of trained head direction cells as the number of head direction cells increases. (c) Top 25 (sorted by directional information rate) Skaggs and Spectral-directional information rate trained neurons from models with 256 output neurons. (d) Histogram of path invariance discrepancy across 1000 trajectories from all untrained models, models trained with Skaggs' directional information, and models trained with spectral-directional information.}
    \label{fig:hd_cell_tuning}
\end{figure}

Using techniques described in Section \ref{sec:SSSE}, we may also train RNNs to encode direction without external targets. Here, the input to the RNN is given by a sequence of turns $\{v_1, \ldots v_{n_x}\}$ where $v_i \in \R^1$ and an initial head direction. Now, similar to what was done for position encoding, we create the \textit{directional} information matrix via the RNN output and calculate epoch loss. Again, we are optimizing the spectral-directional information, $\lambda_1$, or the Skaggs' directional information. While neurons before training fire sporadically, without preferred directions, after training, neurons resemble head-direction cells by firing only when the agent is facing a particular direction (Figure \ref{fig:hd_cell_tuning} (c)). Furthermore, the neurons often fire in non-overlapping directions, as predicted by Theorem \ref{thm:largest_eigenvalue}.

Contrary to place cell learning, spectral-directional information shows no statistically significant improvement over Skaggs' directional information for head direction cell learning (Figure \ref{fig:hd_cell_tuning} (b)). Correspondingly, the distribution of head direction firing fields is similar across models trained with spectral-directional information and Skaggs' directional information, with both techniques producing neurons that fire throughout the stimulus space (Figure \ref{fig:hd_cell_tuning} (a)) and both models are path invariant with respect to permutations in the sequence of turns (Figure \ref{fig:hd_cell_tuning} (d)).

\section{Discussion}
\label{sec:discussion}

To create an internal map of the world, the brain uses a multitude of specialized neurons to encode its environment. Here, we formulated the joint stimulus information rate (Definition \ref{JSI def}) and spectral-spatial information (Definition \ref{stimulus information matrix}). This crucial step allows us to effectively quantify the efficiency of a population of neurons in conveying information about a stimulus. These metrics capture both spatial selectivity and anti-correlation in population codes by explicitly incorporating correlation into the joint stimulus information rate. We applied our metrics to real neural data, demonstrating that they are capable of explaining the differences in encoding efficiency among both pairs and populations of neurons. Finally, we demonstrated that spectral-stimulus information can be used to train RNNs to produce place cell-like and head direction cell-like representations.

Our results show that spectral-spatial information optimization generally produces superior place cell-like encoding than the traditional Skaggs' spatial information rate, as exhibited by larger place cell scores, better decoding performance, less redundancy, and more uniform firing across the spatial domain. However, it is worth noting that for models trained with 256 neurons, spectral-spatial information optimization is, at best, as good as Skaggs' stimulus information optimization it terms of place cell score, although there are still a discrepancies in redundancy and decoding capabilities. This suggests that, as the number of neurons becomes large, role of pairwise correlations in spectral-stimulus information becomes less relevant. Furthermore, despite serving as an excellent metric for place cell population encoding efficiency in both artificial and real neural systems, there is no evidence suggesting that spectral-spatial information is biologically plausible. 

Contrary to previous works, the units in the hidden layer of our models bore little resemblance to that of grid cells. This phenomenon has been observed in supervised models with realistic place cell-like heterogeneous responses [\cite{schaeffer2022no}]. It has been suggested that for place cell-based recurrent architectures to learn grid cell-like hidden layer representations, the readout code must be translation invariant [\cite{banino2018vector, sorscher2019unified}]. This property is not likely present in \textit{in vivo} place cells, which over-represent certain locations, and is not present in those learned by our models [\cite{dupret2010reorganization, gauthier2018dedicated}]. While our models were able to learn many place-cell like responses, some neurons were completely silenced during training (Figure \ref{fig:all_activations}). We hypothesize that this occurred because, in the neural state space, a neuron that fires in an extremely localized region is very close to one that does not fire at all; thus, small changes to the model’s parameters toward the end of training can lead to complete silencing. This behavior is a consequence of optimizing metrics expressed in bits per spike, which reward highly specific responses. While using $I_{sec}$ instead of $I_{spike}$ would mitigate this issue, directly optimizing measures expressed in bits per second could lead to unbounded increases in firing rates.

Our definition of $I_{spike}(A, B:S)$ has two main limitations. First, it is not well-defined for all possible collections of firing rates $\{\lambda_A(x_i)\}_{i = 1}^{n_x}$, $\{\lambda_B(x_i)\}_{i = 1}^{n_x}$, due to potential nonpositive logarithm arguments. We circumvent this by assigning the value $0$ to terms with nonpositive logarithm arguments, as is commonly done for $0\log(0)$ in information-theoretic quantities since $\lim\limits_{x \to 0} x\log(x) = 0$. Secondly, $I_{spike}(A, B:S)$ is not strictly non-negative, even when only considering well-defined terms. However, when the stimulus domain is discretized finely (i.e. $n_s \ge 20$), the joint stimulus information rate is nonnegative in practice (Figure \ref{fig:info_hists}). For self-supervised stimulus encoding (Section \ref{sec:SSSE}), we use $n_s = 100$. Correspondingly, we do not encounter a negative joint stimulus information rate when performing self-supervised stimulus encoding.

A natural extension of our work is to construct stimulus information measures for $n_p \ge 3$ place cells without the use of eigenvalue analysis. However, such measures would likely suffer from the curse of dimensionality, whereas spectral-stimulus information scales quadratically with the number of neurons. Solutions will require further mathematical analysis, possibly diverging from techniques used thus far. Further research could also explore multiobjective loss functions which combine Skaggs' or spectral information rates with additional terms, such as an explicit penalty on certain firing patterns.  

Our results lend themselves nicely to the investigation of place cell remapping. Studies have shown that place cells form statistically independent representations of pairs of environments [\cite{alme2014place}]. Furthermore, evidence suggests that different behavioral strategies can affect the firing of place cells [\cite{levy2023hippocampal, markus1994spatial}]. Here, we assumed the occupancy distribution to be a discrete uniform distribution for all trajectories. Future work on the impact of non-uniform occupancy distributions on place cell firing may prove beneficial. Another avenue for future research is the merging of information-theoretic methods with models of episodic memory in the hippocampus [\cite{benna2021place, squire2004medial}]. Future computational directions could include the implementation of RNNs capable of encoding multiple stimuli at once, a property present in the mammalian brain [\cite{sanders2019temporal}]. This could incorporate an attention mechanism that allows the model to dynamically decide how much to focus on various stimuli. Recent interest has also been in using information theory to understand the phenomenon in which neural networks during training suddenly generalize, known as grokking [\cite{clauw2024information}] and image segmentation [\cite{savarese2021information}].  

Overall, our work displays the success of unifying information theory with neuroscience and its applications to machine learning. The result is a widely applicable formula for neuron population encoding efficiency, which lays the foundation for more questions regarding the formation, function, and role of specialized neurons in the mammalian brain.  

\section*{Acknowledgments}
This work has been partially supported by the Army Research Laboratory Cooperative Agreement No W911NF2120186, and STRONG ARL CA No W911NF-22-2-0139.

\begin{supplement}
\stitle{Supplement A}
\sdescription{Experimental details, proofs, and sensitivity analysis.}
\end{supplement}
\begin{supplement}
\stitle{Supplement B}
\sdescription{Additional place cell activation maps.}
\end{supplement}
\begin{supplement}
\stitle{Supplement C}
\sdescription{Source code for computing information theoretic measures and performing self-supervised stimulus encoding.}
\end{supplement}

%%  MR numbers will be added by VTeX.                      %%
%%                                                         %%
%%  Use \cite{...} to cite references in text.             %%
%%                                                         %%
%%%%%%%%%%%%%%%%%%%%%%%%%%%%%%%%%%%%%%%%%%%%%%%%%%%%%%%%%%%%%

%% if your bibliography is in bibtex format, uncomment commands:
%\bibliographystyle{imsart-number} % Style BST file (imsart-number.bst or imsart-nameyear.bst)
%\bibliography{bibliography}       % Bibliography file (usually '*.bib')

%% or include bibliography directly:
\bibliographystyle{imsart-nameyear}
\bibliography{references}

\renewcommand{\thefigure}{S\arabic{figure}}
\renewcommand{\thetable}{S\arabic{table}}
\renewcommand{\thesection}{S\arabic{section}}
\setcounter{figure}{0}
\setcounter{table}{0}
\setcounter{section}{0}
\begin{frontmatter}
\title{Supplementary Materials for Spectral-Stimulus Information for Self-Supervised
Stimulus Encoding}
%\title{A sample article title with some additional note\thanksref{t1}}
\runtitle{Spectral-Stimulus Information for Self-Supervised
Stimulus Encoding}
%\thankstext{T1}{A sample additional note to the title.}

\begin{aug}
%%%%%%%%%%%%%%%%%%%%%%%%%%%%%%%%%%%%%%%%%%%%%%%
%% Only one address is permitted per author. %%
%% Only division, organization and e-mail is %%
%% included in the address.                  %%
%% Additional information can be included in %%
\mycomment{
\author[A]{\fnms{Jared}~\snm{Deighton} \ead[label=e6]{jdeighto@vols.utk.edu}},
\author[B]{\fnms{Wyatt}~\snm{Mackey}\ead[label=e7]{wyatt.t.mackey.civ@army.mil}},
\author[B]{\fnms{Ioannis}~\snm{Schizas}\ead[label=e8]{ioannis.d.schizas.civ@army.mil}},
\author[B]{\fnms{David L.}~\snm{Boothe}\ead[label=e9]{david.l.boothe7.civ@army.mil}}
\and
\author[A]{\fnms{Vasileios}~\snm{Maroulas}\thanks{Corresponding author}\ead[label=e10]{vmaroula@utk.edu}}

%%%%%%%%%%%%%%%%%%%%%%%%%%%%%%%%%%%%%%%%%%%%%%
%% Addresses                                %%
%%%%%%%%%%%%%%%%%%%%%%%%%%%%%%%%%%%%%%%%%%%%%%
\address[A]{Department of Mathematics, The University of Tennessee, Knoxville, TN 37996, United States of America \printead[presep={,\ }]{e6,e10}}

\address[B]{The U.S. Army Combat Capabilities Development Command Army Research Laboratory, Aberdeen Proving Ground, MD 21005, United States of America \printead[presep={,\ }]{e7,e8,e9}}
}
\end{aug}

\mycomment{
\begin{keyword}[class=MSC]
\kwd[Primary ]{92B20}
\kwd{94A15}
\kwd[; secondary ]{92-10}
\end{keyword}

\begin{keyword}
\kwd{Information Theory}
\kwd{Recurrent Neural Networks}
\kwd{Place Cells}
\kwd{Neural Coding}
\end{keyword}
}

\end{frontmatter}
%%%%%%%%%%%%%%%%%%%%%%%%%%%%%%%%%%%%%%%%%%%%%%
%% Please use \tableofcontents for articles %%
%% with 50 pages and more                   %%
%%%%%%%%%%%%%%%%%%%%%%%%%%%%%%%%%%%%%%%%%%%%%%
%\tableofcontents
\section{Experimental Details}\label{experimental_details}

Our code was implemented in PyTorch [\cite{paszke2019pytorch}] and is available at \url{https://github.com/JaredDeightonUTK/Information-PC-Paper}.
Our methodology closely aligns with the frameworks established in the publicly available codes used in \cite{sorscher2019unified, sorscher2023unified} and in \cite{cueva2018emergence}. Hyperparameters for our
experiments are listed in Table \ref{table:hyperparameters}. The majority of the code was run using a 2.8 GHz Quad-Core Intel Core i7 processor with 16 GB of memory. Here, training a single spectral-spatial information model with 16 place cells and 256 hidden units for 100 epochs takes roughly $16$ seconds. Training 64 place cells and 1028 hidden units for 30 epochs takes roughly 2 minutes. Code to train models with 256 place cells and to perform the sensitivity analysis was run using a Tesla K80 GPU with 64GB of memory.

\begin{table}[h]
\centering
\caption{Hyperparameters used for the RNNs.}
\label{table:hyperparameters}
\begin{tabular}{ll}
\toprule
\textbf{Hyperparameter}        & \textbf{Value}               \\ \hline
Batch size   & 40                            \\
Sequence length  & 100                       \\
Learning rate & 1e-4                         \\
Number of output units  & 16, 32, 64, 128, 256                           \\
Number of hidden units  & 256, 1028, 1028, 1028, 2048                         \\
Periodic boundary conditions & False                \\
Arena Size (m) & 0.5                  \\
RNN nonlinearity  & $ReLU(\cdot)$                        \\
Optimizer & Adam [\cite{kingma2014adam}]
\\
Weight decay & None \\
\bottomrule
\end{tabular}
\end{table}

\mycomment{
\begin{table}[h]
\centering
\caption{Hyperparameters used for the FFNN.}
\label{table:hyperparameters_FFNN}
\begin{tabular}{ll}
\toprule
\textbf{Hyperparameter}        & \textbf{Value}               \\ \hline
Batch size   & 32                            \\
Learning rate & 1e-3                         \\
Number of layers & 2                  \\
Hidden Layer Sizes  & 128, 128                      \\
Dropout & 0.2                \\
FFNN nonlinearity  & $ReLU(\cdot)$                        \\
Optimizer & Adam [\cite{kingma2014adam}]
\\
Weight decay & None \\
\bottomrule
\end{tabular}
\end{table}}

\begin{figure}[!htb]
    \centering
    \includegraphics[width = 0.9\linewidth]{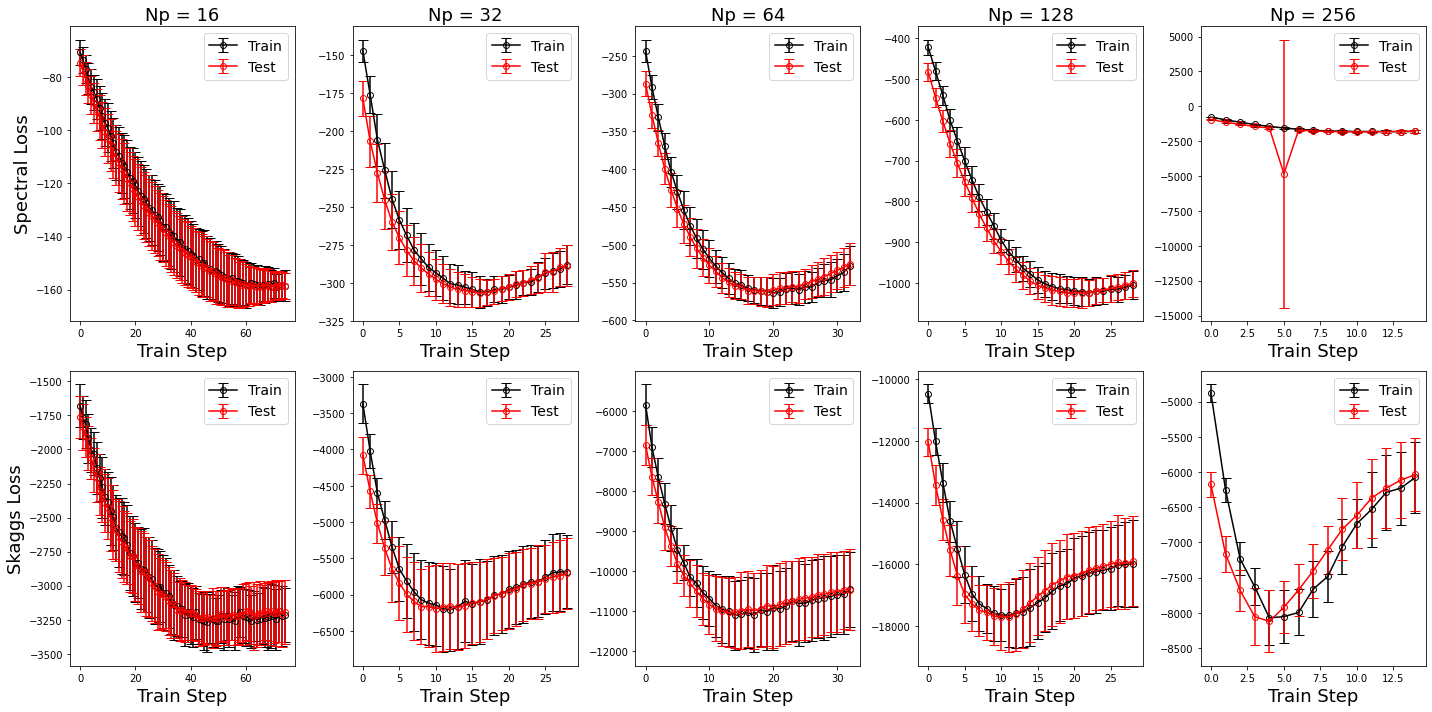}
    \caption{Mean ($\pm$ standard deviation) train and test loss from 10 experiments for each $N_p$ (number of place cells) and loss function.}
    \label{fig: validation lsos}
\end{figure}

\subsection{Trajectory generation} We generate trajectories in a two-dimensional square arena using previously published codes found in  \cite{sorscher2019unified, sorscher2023unified}. In short, the agent's initial position is randomly drawn from a uniform distribution across the set of possible positions. Then, a sequence of random speeds and turns is constructed to create the trajectory. Trajectories can be freely generated. Training and testing summaries are in Figure \ref{fig: validation lsos}. 

\subsection{Initial Activation}
The initial neuron activations $P_0$ are generated using a difference of softmaxes. First, we randomly choose $n_p$ two-dimensional points in the arena, denoted as $y \in \R^{n_p \times 2}$. Then, for a position $x \in \R^2$, we have 
\begin{align*}
    (P_0)_i = \text{softmax}\left(\frac{-||x - y_i||^2}{2\sigma_E^2}\right) - \text{softmax}\left(\frac{-||x - y_i||^2}{2\sigma_I^2}\right) 
\end{align*}
for $i = 1,2, \ldots, n_p$. $P_0$ is then shifted and scaled such that outputs so that they lie in [0,1].

\subsection{Leave-One-out classification} Leave-one-out classification for spatial decoding was performed using Scikit-learn's [\cite{scikit-learn}] Naive Bayes classifier and KBinsDiscretizer. For each trained model, we pass 10 trajectories (1000 positions) through the network to obtain place cell activations. The 1000 positions are binned using the KBinsDiscretizer with 100 bins in each dimension. Then, all but one of the firing rate-location pairs are used to estimate posterior probabilities (i.e. to fit the classifier). Then, the Naive Bayes classifier predicts the position of the left-out firing rate. This process is repeated so that every firing rate value is used for testing. The predicted positions (given as a predicted bin) are then transformed back to a predicted position in $\R^2$ using KBinsDiscretizer's inverse transform, from which the mean-squared error with the true positions is calculated.

\subsection{Poisson-Bayes classification} 
We implemented a Bayesian decoder based on a generative Poisson model of neural firing. For each trained network, place cell firing rates were used to simulate Poisson spike trains with a time step of one second. The corresponding positions were discretized with 50 bins in each dimension. Eighty percent of the data were used to estimate tuning curves, defined as the average firing rate of each neuron within each spatial bin. These tuning curves specify, for each position, the expected Poisson firing rate of every neuron. For the remaining twenty percent of the data, decoding was performed by evaluating the likelihood of the observed spike vector at each spatial bin under the Poisson model. Assuming a uniform prior over space, these likelihoods were normalized to form posterior probabilities over position. The decoded position at each time step was taken as the spatial bin with the maximum posterior probability, which was then mapped back to a physical coordinate in $\mathbb{R}^2$. 

\mycomment{
\subsection{Support Vector Machine.} Quadrant classification for spatial decoding was performed using Scikit-learn's [\cite{scikit-learn}] support vector machine (SVM) and KBinsDiscretizer.  First, we generate many trajectories (1000) to obtain place cell activations at various positions for each model. Now, to create the quadrant classes the positions are binned using the KBinsDiscretizer with 2 bins in each dimension. We use 10000 firing rate-quadrant pairs to fit an SVM with a linear kernel. Then, the SVM is tested on 1000 separate samples to obtain the quadrant classification accuracy.}

\subsection{Statistical Analysis} To assess the differences between the neural representation metrics derived from two training objectives, spectral and Skaggs, we used independent two-sample t-tests for each metric across a range of model sizes using the SciPy [\cite{2020SciPy-NMeth}] stats package. 

\subsection{Neural Recordings}

\subsubsection{Head direction cells}\label{subsub:HD_cells}

We look at a subset of head direction cell spiking data collected by Peyrache et al.~in [\cite{peyrache2015hd}]. Specifically, we analyze 20 minutes of spiking data collected from `Mouse28-140313' during the foraging portion of the experiment. During this time, the mouse was allowed to roam around its rectangular environment. 
Researchers recorded the mouse's ground truth head direction and spiking data from 22 head direction cells within the anterodorsal thalamic nucleus (ADn) binned into $100$ms-wide bins. 

\subsubsection{Place cells}\label{subsub:PCs}

Place cells are neurons in the hippocampus that fire when an animal visits specific regions of its environment [\cite{o1978hippocampus, tolman1948cognitive}]. We apply our information-theoretic measures to place cells recording collected by Hazama et al.~in [\cite{hazama2019data}] and Harland et al.~in [\cite{harland2021dorsal}]. From Hazama et al., spike-timestamps were recorded of 18 place cells located in the right hippocampal CA1 subfield of a male Japanese monkey performing a shuttling task on a rectangular ($4 \times 0.9 \ m$) track. The monkey was required to shuttle back and forth on the track multiple times to get rewards. From Harland et al., rats were recorded in a large open environment (“megaspace,”($18.6 \ m^2$) and a smaller classic environment ($2.2 \ m^2$) to compare place cell properties across scales. Using wireless recording and a food-baited robot to ensure full spatial coverage, the study found that place cells in the megaspace exhibited multiple spatially distributed subfields of varying sizes, with average field size increasing in larger environments. The number of subfields per cell did not correlate with the total area covered, and robot-following behavior produced comparable spatial coding to traditional foraging.

\section{Place cell score}\label{app:pc_score}
The place cell score is a combination of three characterizing features of place cells: (i) smoothness of the firing rate; (ii) binary firing rates across space; and (iii) sparse activations. Each of these factors is a key feature of \emph{in vitro} place cells [\cite{bures1997place, markus1994spatial}]. The place cell score, $\operatorname{PC}$, of a place cell $A$ is given by
\begin{gather}
    \label{place cell score}
    \text{PC}_{\text{score}}(A, \epsilon) = 10*\text{Binary}(A, \epsilon)- 100*\text{Smoothness}(A) -10*\text{Sparsity}(A),
\end{gather}
where $\epsilon$ is a small number and a hyperparameter of the system. To compute the smoothness, binary activity, and sparsity scores of a place cell, we measure the firing rates of the place cell $A$ across the entire arena and normalize values between 0 and 1. 
Then, if $\Pi \in \R^{h \times h} $ is the binned matrix of normalized firing rates  across a square arena from place cell $A$ at a spatial resolution $h$, we use the following metrics: 
\begin{align}
    &\text{Smoothness}(A) = \frac{1}{h^2} \sum_{i = 1}^{h} \sum_{j = 1}^{h} |\Pi_{i,j} - \Tilde{\Pi}_{i,j}| \label{smoothness}, \\ 
    &\text{Binary}(A, \epsilon) = P(\Pi < \epsilon) + P(\Pi > 1- \epsilon) \label{binary},  \\  
    &\text{Sparsity}(A) =  \frac{ (\sum_{i = 1}^{h} \sum_{j = 1}^{h} \Pi_{i,j})^2}{\sum_{i = 1}^{h} \sum_{j = 1}^{h} \Pi^2_{i,j}} \label{sparsity}, 
\end{align}
where $\Tilde{\Pi} \in \R^{h \times h}$ is the smoothed matrix of normalized firing rates from place cell $A$ and $P(\Pi < \epsilon)$ is the fraction of entries in $\Pi$ that are less than $\epsilon$. For our purposes, $\Tilde{\Pi}$ is acquired via Gaussian Blur with kernel size $(5, 5)$ and kernel standard deviations $\sigma_x = \sigma_y = 1$. We may also calculate the spatial information rate of the learned place cells across the entire arena via 
\begin{gather}
\label{arena spatial information}
    I_{spike}(A) = \sum_{i = 1}^{h} \sum_{j = 1}^{h} \frac{\Pi_{i,j}}{\bar{\Pi}} \log_2\left(\frac{\Pi_{i,j}}{\bar{\Pi}}\right)p(\text{bin}_{i,j}),
\end{gather}
where $\bar{\Pi} =  \sum_{i = 1}^{h} \sum_{j = 1}^{h}p(\text{bin}_{i,j}) \Pi_{i,j}$. For simplicity, we assume the agent is equally likely to be in any spatial bin, i.e. $p(\text{bin}_{i,j}) = 1/h^2$ for all $i,j$.

\section{Derivations and Proofs}\label{app:proofs}

\begin{table}[!htbp]
\centering
\caption{Mathematical Notation}
\label{table:mathematical notation}
\begin{tabular}{ll}
\toprule
\textbf{Notation}        & \textbf{Value}               \\ \hline
$J$  & Stimulus information matrix\\
$P_t$  & Output of RNN at step $t$\\
$G_t$  & Hidden state of RNN at step $t$\\
$W_{ih}$ & RNN input-hidden matrix\\
$W_{hh}$ & RNN hidden-hidden update matrix\\
$\lambda_A(s)$ & Firing rate of neuron $A$ when stimulus $s$ is present\\
$\bar{\lambda}_A$ & Overall mean firing rate of neuron $A$ \\
$\lambda_1$ & Spectral Stimulus Information\\
$r$ & Pearson's correlation coefficient\\
$x^T$ & The tanspose of the vector $x$ \\
\bottomrule
\end{tabular}
\end{table}

\begin{const}[Derivation of the Joint Stimulus Information Rate]
\label{const JSIR}

Here, we construct the joint stimulus information rate of two neurons $A$ and $B$.  For some stimulus space $S$, let $\lambda_A(s)$ and $\lambda_B(s)$ be the firing rate of neurons $A$ and $B$ at location $s \in S$, respectively. Let $A_s$ and $B_s$ denote the events where $A$ and $B$ spike at stimulus $s$ during $\Delta t$ and $A_s^c$ and $B_s^c$ denote the complements. 

Now, we estimate the correlation between two neurons $r$ as the Pearson's correlation over the entire spatial domain and create the joint distribution of the cells firing during a short time $\Delta t$ by treating each distribution of a place cell firing during $\Delta t$ as a Bernoulli distribution. Then, 
$$R = COV(A,B) = r \sqrt{\lambda_A(s)\Delta t (1 - \lambda_A(s)\Delta t)\lambda_B(s)\Delta t(1 - \lambda_B(s)\Delta t)}$$ where $r$ is the correlation coefficient between neurons $A$ and $B$, and we have
\begin{align*}
    &P(A_s \cap B_s) = R + \lambda_A(s) \lambda_B(s) (\Delta t)^2\\
    &P(A_s \cap B_s^c) = \lambda_A(s)\Delta t(1 - \lambda_B(s)\Delta t) - R \\
    &P(A_s^c \cap B_s) =(1 - \lambda_A(s)\Delta t)\lambda_B(s)\Delta t - R \\
    &P(A_s^c\cap B_s^c) = (1 - \lambda_A(s)\Delta t)(1 - \lambda_B(s)\Delta t) + R
\end{align*}
Neglecting terms higher than $\Delta t$, we have $R \approx r\sqrt{\lambda_A(s) \lambda_B(s)} \Delta t$ and
\begin{align*}
    &P(A_s \cap B_s) = r\sqrt{\lambda_A(s) \lambda_B(s)} \Delta t \\
    &P(A_s \cap B_s^c) = \lambda_A(s)\Delta t - r\sqrt{\lambda_A(s) \lambda_B(s)} \Delta t \\
    &P(A_s^c \cap B_s) = \lambda_B(s)\Delta t - r\sqrt{\lambda_A(s) \lambda_B(s)} \Delta t \\
    &P(A_s^c \cap B_s^c)= 1 - \lambda_A(s)\Delta t- \lambda_B(s)\Delta t + r\sqrt{\lambda_A(s) \lambda_B(s)} \Delta t
\end{align*}

We also have, regardless of location $E[(P(A_s \cap B_s))] = \sum_s p(s)r\sqrt{\lambda_A(s) \lambda_B(s)} \Delta t$ and similarly for the other scenarios. 

Then, the mutual information between the joint distribution and space is
\begin{align}
\label{Initial Information Eq}
    I(A,B) &= \sum_s \sum_{\omega_s \in \Omega_s}p(s) P(\omega_s) \log_2\left(\frac{ P(\omega_s)}{ E[P(\omega_s)]}\right)
\end{align} 
where $\Omega_s = \{A_s \cap B_s, \ldots, A_s^c \cap B_s^c \}$ is the set of possible outcomes at location $s$. 
In Lemma \ref{no-fire lemma}, we show that the final term ($\omega_s = A_s^c \cap B_s^c$) goes to zero when dropping terms higher than $\Delta t$. Thus, expanding Equation \eqref{Initial Information Eq} and a short simplification yields, 
\begin{align*}
    I(A,B) &= \Delta t \sum_s p(s) r\sqrt{\lambda_A(s) \lambda_B(s)} \log_2\left(\frac{\sqrt{\lambda_A(s) \lambda_B(s)} }{\sum_s p(s)\sqrt{\lambda_A(s) \lambda_B(s)}}\right) \\
    &+ p(s)[\lambda_A(s) - r\sqrt{\lambda_A(s) \lambda_B(s)} ]\log_2\left(\frac{\lambda_A(s)- r\sqrt{\lambda_A(s) \lambda_B(s)} }{\sum_s p(s)[ \lambda_A(s) - r\sqrt{\lambda_A(s) \lambda_B(s)} ]} \right) \\
    &+ p(s) [\lambda_B(s) - r\sqrt{\lambda_A(s) \lambda_B(s)} ]\log_2\left(\frac{\lambda_B(s) - r\sqrt{\lambda_A(s) \lambda_B(s)} }{\sum_s p(s)[\lambda_B(s)- r\sqrt{\lambda_A(s) \lambda_B(s)} ]} \right)
\end{align*}

Then, using the series expansion of $I(A,B)$ around $t$, we have that the first time derivative is approximated by
\begin{align*}
    I(A,B)' &\approx \sum_s p(s) r\sqrt{\lambda_A(s) \lambda_B(s)} \log_2\left(\frac{\sqrt{\lambda_A(s) \lambda_B(s)} }{\sum_s p(s)\sqrt{\lambda_A(s) \lambda_B(s)}}\right) \\
    &+ p(s)[\lambda_A(s) - r\sqrt{\lambda_A(s) \lambda_B(s)} ]\log_2\left(\frac{\lambda_A(s)- r\sqrt{\lambda_A(s) \lambda_B(s)} }{\sum_s p(s)[ \lambda_A(s) - r\sqrt{\lambda_A(s) \lambda_B(s)} ]} \right) \\
    &+ p(s) [\lambda_B(s) - r\sqrt{\lambda_A(s) \lambda_B(s)} ]\log_2\left(\frac{\lambda_B(s) - r\sqrt{\lambda_A(s) \lambda_B(s)} }{\sum_s p(s)[\lambda_B(s)- r\sqrt{\lambda_A(s) \lambda_B(s)} ]} \right)
\end{align*}
This is the definition of $I_{sec}(A,B)$ the joint spatial information rate. We may write more succinctly, 
\begin{align*}
    I_{sec}(A,B) &= \sum_s p(s) r\sqrt{\lambda_{A,B}(s)}\log_2\left(\frac{\sqrt{\lambda_{A,B}(s)}}{\tilde{\lambda}_{A,B}} \right)\\ 
    &+ p(s)\left(\lambda_A(s) - r\sqrt{\lambda_{A,B}(s)}\right)\log_2\left(\frac{\lambda_A(s) - r\sqrt{\lambda_{A,B}(s)}}{\bar{\lambda}_A - r\tilde{\lambda}_{A,B}}\right)\\
    &+ p(s)\left(\lambda_B(s) - r\sqrt{\lambda_{A,B}(s)}\right)\log_2\left(\frac{\lambda_B(s) - r\sqrt{\lambda_{A,B}(s)}}{\bar{\lambda}_B - r\tilde{\lambda}_{A,B}}\right)
\end{align*}
where $\sum_s p(s) \lambda_A(s) = \Bar{\lambda}_A$, $\sum_s p(s) \lambda_B(s) = \Bar{\lambda}_B$, $\sum_s p(s) \sqrt{\lambda_{A,B}}(s) = \tilde{\lambda}_{A,B}$, and $\lambda_A(s)\lambda_B(s) = \lambda_{A,B}(s)$. 
\mycomment{
Notice, if $A,B$ are independent ($r = 0$), then, 
\begin{align*}
    I_{sec}(A,B) &= \sum_s p(s) 0 + p(s)\left(\lambda_A(s) - 0\right)\log_2\left(\frac{\lambda_A(s) - 0}{\bar{\lambda}_A  - 0}\right)
    + p(s)\left(\lambda_B(s) - 0\right)\log_2\left(\frac{\lambda_B(s) - 0}{\bar{\lambda}_B - 0}\right)\\
     &= \sum_s p(s)\lambda_A(s)\log_2\left( \frac{\lambda_A(s)}{\bar{\lambda_A}}\right) + \sum_s p(s)\lambda_B(s)\log_2\left( \frac{\lambda_B(s)}{\bar{\lambda_B}}\right)\\
     &= I_{sec}(A) + I_{sec}(B)
\end{align*}
as expected.}

Finally as is done for $I_{sec}$, we may write the joint information of in units of bits/spike by defining,
\begin{gather*}
    I_{spike}(A,B) = \frac{1}{\Lambda_{A, B}} I_{sec}(A,B)
\end{gather*}
where $ \lambda_{A,B} = \sum_s p(s) \frac{\lambda_A(s) + \lambda_B(s)}{2} = \frac{\Bar{\lambda}_A + \Bar{\lambda}_B}{2}$. This is chosen so that $I_{spike}(A,A) = I_{spike}(A)$. 

\end{const}

\begin{lemma}[No-firing term provides no information] 
\label{no-fire lemma}
Here we show that the no-firing term from Equation \eqref{Initial Information Eq} goes to zero. 

\begin{proof}
    First, notice
\begin{align*}
    &\log_2\left(\frac{1 - \Delta t(\lambda_A(s) + \lambda_B(s) - r\sqrt{\lambda_A(s) \lambda_B(s)} )}{\sum_s p(s) [1 - \Delta t(\lambda_A(s) + \lambda_B(s) - r\sqrt{\lambda_A(s) \lambda_B(s)} )}\right) \\
    &= \log_2\left(1 - \Delta t(\lambda_A(s) + \lambda_B(s) - r\sqrt{\lambda_A(s) \lambda_B(s)} \right)\\
    &- \log_2\left(\sum_s p(s)  - \Delta t \sum_s p(s) [\lambda_A(s) + \lambda_B(s) - r\sqrt{\lambda_A(s) \lambda_B(s)}] \right)\\
    &= \log_2\left(1 - \Delta t(\lambda_A(s) + \lambda_B(s) - r\sqrt{\lambda_A(s) \lambda_B(s))} \right)\\
    &- \log_2\left(1  - \Delta t \sum_s p(s) [\lambda_A(s) + \lambda_B(s) - r\sqrt{\lambda_A(s) \lambda_B(s)}] \right)
\end{align*}

Using the Taylor series $\log_2(1 - x\Delta t) \approx -\frac{x\Delta t}{\ln(2)}$ we can approximate these terms by
\begin{align*}
    &\log_2\left(\frac{1 - \Delta t(\lambda_A(s) + \lambda_B(s) - r\sqrt{\lambda_A(s) \lambda_B(s)} )}{\sum_s p(s) [1 - \Delta t(\lambda_A(s) + \lambda_B(s) - r\sqrt{\lambda_A(s) \lambda_B(s)} )}\right) \\
    &\approx -\Delta t\frac{\lambda_A(s) + \lambda_B(s) - r\sqrt{\lambda_A(s) \lambda_B(s)}}{\ln(2)} + \Delta t \frac{\sum_s p(s) [\lambda_A(s) + \lambda_B(s) - r\sqrt{\lambda_A(s) \lambda_B(s)} ]}{\ln(2)}
\end{align*}
Thus, focusing on the last term of  $I(A,B)$ in Equation \eqref{Initial Information Eq}, 
\begin{align*}
    &\sum_s p(s) [1 - \Delta t(\lambda_A(s) + \lambda_B(s) - r\sqrt{\lambda_A(s) \lambda_B(s)} )]\\
    &\cdot \log_2\left(\frac{1 - \Delta t(\lambda_A(s) + \lambda_B(s) - r\sqrt{\lambda_A(s) \lambda_B(s)} )}{\sum_s p(s) [1 - \Delta t(\lambda_A(s) + \lambda_B(s) - r\sqrt{\lambda_A(s) \lambda_B(s)} )]}\right)\\
    &= \sum_s p(s) \frac{\Delta t}{\ln(2)} \left(\sum_sp(s)[\lambda_A(s) + \lambda_B(s) - r\sqrt{\lambda_A(s)\lambda_B(s)}] - (\lambda_A(s) + \lambda_B(s) - r\sqrt{\lambda_A(s)\lambda_B(s)})] \right)\\
    &= 0
\end{align*}
\end{proof}
\end{lemma}

\mycomment{
\begin{rem}[Limit Argument]
\label{Limit Argument}
Here, we show that for Equations \eqref{I_sec(A,B)} and \eqref{I_spike(A,B)}, the quotients involving $0\log_2(0/0)$ may be defined as 0. 

There are three cases for $I_{sec}$ in which $0\log_2(0/0)$ may arise.

\textit{Case 1:} First, suppose that $\Tilde{\lambda}_{A,B} = 0$ for some neurons $A$ and $B$. Then, for any $x \in X$, either $\lambda_{a}(x) = 0$ or $\lambda_B(x)  = 0$. Either way, $\sqrt{\lambda_{A,B}(x)} = 0$. Thus, this scenario only applies to the trivial case when both neurons have no firing field. Here, of course, the information is zero. 

\textit{Case 2:} Next suppose that $\Bar{\lambda}_A(x) - r \Tilde{\lambda}_{A,B} = 0$. It is easy to check that $\Bar{\lambda}_A(x) - r \Tilde{\lambda}_{A,B} = 0$ if and only if $\lambda_A(x) = \lambda_B(x)$ for all $x \in X$. This is the scenario in which $r = 1$, and $\lambda(a) - r \sqrt{\lambda_{A,B}(x) = 0}$ for all $x \in X$. Thus, $0\log_2(0/0)$ only arises when the neurons have identical firing fields. Here, the information accumulated from any term of Equation \eqref{I_sec(A,B)} besides the first should be zero.

\textit{Case 3:} An similar analysis of \textit{Case 2} holds for when $\Bar{\lambda}_B(x) - r \Tilde{\lambda}_{A,B} = 0$. 
\end{rem}

}

\begin{prop}[Stimulus Information Properties]
\label{properties} We show that the described properties hold. First, $I_{spike}(A,B) = I_{spike}(B,A)$ holds trivially. Next, we show that $I_{spike}(A,A) = I_{spike}(A)$
\begin{proof}
    
Notice that if $A = B$, then we have the following:
\begin{align*}
    &\Bar{\lambda}_B = \Bar{\lambda}_A\\
    &\sqrt{\lambda_{A,B}(x)} = \sqrt{\lambda_{a}(x)\lambda_{b}(x)} = \sqrt{\lambda^2_{a}(x)} = \lambda_A(x)\\
    &\Tilde{\lambda}_{A,B} = \sum_x p(x)\sqrt{\lambda_{A,B}(x)} = \sum_x p(x)\lambda_{a(x)} = \Bar{\lambda}_A \\
    &\lambda_{A,B} = \frac{\Bar{\lambda}_A + \Bar{\lambda_B}}{2} = \frac{\Bar{\lambda}_A + \Bar{\lambda_A}}{2} = \Bar{\lambda}_A\\
    &r = 1
\end{align*}
Thus, 
\begin{align*}
    I_{spike}(A,A) &= \frac{1}{\Bar{\lambda}_A}I_{sec}(A,A)
    = \frac{1}{\Bar{\lambda}_A}\sum_{s \in S} \Bigg[ p(s) \lambda_{a}(s)\log_2\left(\frac{\lambda_{a}(s)}{\Bar{\lambda}_A} \right) \\ &+ p(s)\left(\lambda_A(s) - \lambda_A(s)\right)\log_2\left(\frac{\lambda_A(s) - \lambda_A(s)}{\bar{\lambda}_A - \bar{\lambda}_A}\right)  \\
    &+ p(s)\left(\lambda_A(s) - \lambda_A(s)\right)\log_2\left(\frac{\lambda_A(s) - \lambda_A(s)}{\bar{\lambda}_A - \bar{\lambda}_A}\right)  \Bigg]\\
    &= \sum_{s \in S} \Bigg[ p(s)\frac{\lambda_{a}}{\Bar{\lambda}_A}(s)\log_2\left(\frac{\lambda_{a}(s)}{\Bar{\lambda}_A} \right) + p(s)0\log(0/0) + p(s)0\log(0/0)\Bigg]
\end{align*}
Lastly, setting terms which are not well-defined to zero, 
\begin{gather*}
     I_{spike}(A,A) = \sum_{s \in S} p(s)\frac{\lambda_{a}}{\Bar{\lambda}_A}(s)\log_2\left(\frac{\lambda_{a}(s)}{\Bar{\lambda}_A} \right)  = I_{spike}(A)
\end{gather*}
\end{proof}
\end{prop}

\mycomment{
\begin{const}
\label{stim_info_deriv}
Consider
\begin{gather*}
    I_{spike}(S) = \sum_{s \in S} \frac{\lambda(s)}{\bar{\lambda}} \log_2 \left(\frac{\lambda(s)}{\bar{\lambda}}\right) p(s),
\end{gather*}
where $\bar{\lambda} = \sum_{s \in S}\lambda(s) p(s)$. We find the derivative of $ I_{spike}(S)$ with respect to a single $\lambda(s')$. 

\begin{gather*}
    \frac{\partial I(S)}{\partial\lambda(s')} = \sum_{s \in S} \frac{\partial}{\partial\lambda(s')} \frac{\lambda(s)}{\bar{\lambda}} \log_2 \left(\frac{\lambda(s)}{\bar{\lambda}}\right) p(s)
\end{gather*}
Let $f(s) = \frac{\lambda(s)}{\bar{\lambda}} \log_2 \left(\frac{\lambda(s)}{\bar{\lambda}}\right) p(s)$, then $\frac{\partial I(S)}{\partial\lambda(s')} = \sum_{s \in S} \frac{\partial f(s)}{\partial\lambda(s')}$. We consider two cases, when $s = s'$ and when $s \neq s'$. 

\noindent \textbf{Case 1:} Suppose $s = s'$. Then, 
\begin{gather*}
    \frac{\partial f(s)}{\partial\lambda(s')} = \frac{\partial f(s')}{\partial\lambda(s')} \\
    = \frac{\partial}{\partial\lambda(s')} \left[\frac{\lambda(s')}{\bar{\lambda}} \log_2 \left(\frac{\lambda(s')}{\bar{\lambda}}\right) p(s') \right]\\
    = p(s')\left[\frac{\partial}{\partial\lambda(s')}\left(\frac{\lambda(s')}{\bar{\lambda}}\right)\log_2 \left(\frac{\lambda(s')}{\bar{\lambda}}\right) + \left(\frac{\lambda(s')}{\bar{\lambda}}\right)\frac{\partial}{\partial\lambda(s')}\log_2 \left(\frac{\lambda(s')}{\bar{\lambda}}\right) \right]
\end{gather*}
Now, 
\begin{gather*}
    \frac{\partial}{\partial\lambda(s')}\left(\frac{\lambda(s')}{\bar{\lambda}}\right) = \frac{\bar{\lambda} - \lambda(s')\frac{\partial \bar{\lambda}}{\partial\lambda(s')}}{\bar{\lambda}^2} = \frac{\bar{\lambda} - \lambda(s')p(s')}{\bar{\lambda}^2}
\end{gather*}
And,
\begin{gather*}
    \frac{\partial}{\partial\lambda(s')}\log_2 \left(\frac{\lambda(s')}{\bar{\lambda}}\right) = \frac{\bar{\lambda}}{\ln(2) \lambda(s')}\frac{\partial}{\partial\lambda(s')}\left(\frac{\lambda(s')}{\bar{\lambda}}\right)\\
    = \frac{\bar{\lambda}}{\ln(2) \lambda(s')}\frac{\bar{\lambda} - \lambda(s')p(s')}{\bar{\lambda}^2}
\end{gather*}
Thus,
\begin{gather*}
    \frac{\partial f(s)}{\partial\lambda(s')} = p(s')\left[\frac{\bar{\lambda} - \lambda(s')p(s')}{\bar{\lambda}^2}\log_2 \left(\frac{\lambda(s')}{\bar{\lambda}}\right) + \left(\frac{\lambda(s')}{\bar{\lambda}}\right)\frac{\bar{\lambda}}{\ln(2) \lambda(s')}\frac{\bar{\lambda} - \lambda(s')p(s')}{\bar{\lambda}^2}\right]\\
    = p(s')\left[\frac{\bar{\lambda} - \lambda(s')p(s')}{\bar{\lambda}^2}\log_2 \left(\frac{\lambda(s')}{\bar{\lambda}}\right) + \frac{\bar{\lambda} - \lambda(s')p(s')}{\ln(2)\bar{\lambda}^2}\right]\\
    = p(s')\frac{\bar{\lambda} - \lambda(s')p(s')}{\bar{\lambda}^2}\left[\log_2 \left(\frac{\lambda(s')}{\bar{\lambda}}\right) + \frac{1}{\ln(2)}\right]
\end{gather*}
\noindent \textbf{Case 2:} Suppose $s \neq s'$. Then,
\begin{gather*}
    \frac{\partial f(s)}{\partial\lambda(s')} = \frac{\partial}{\partial \bar{\lambda}} \left[\frac{\lambda(s)}{\bar{\lambda}} \log_2 \left(\frac{\lambda(s)}{\bar{\lambda}}\right) p(s)\right]\frac{\partial \bar{\lambda}}{\partial\lambda(s')}\\
    = p(s)p(s')\left[-\frac{\lambda(s)}{\bar{\lambda}^2}\log_2 \left(\frac{\lambda(s)}{\bar{\lambda}}\right) -\frac{\lambda(s)}{\bar{\lambda}}\frac{1}{\bar{\lambda}}\frac{1}{\ln(2)}\right]\\
    = -\frac{\lambda(s)p(s)p(s')}{\bar{\lambda}^2}\left[\log_2 \left(\frac{\lambda(s)}{\bar{\lambda}}\right) + \frac{1}{\ln(2)}\right]
\end{gather*}
Thus,
\begin{gather*}
     \frac{\partial I(S)}{\partial\lambda(s')} = p(s')\frac{\bar{\lambda} - \lambda(s')p(s')}{\bar{\lambda}^2}\left[\log_2 \left(\frac{\lambda(s')}{\bar{\lambda}}\right) + \frac{1}{\ln(2)}\right]\\ - \sum_{s \in S, s \neq s'}\frac{\lambda(s)p(s)p(s')}{\bar{\lambda}^2}\left[\log_2 \left(\frac{\lambda(s)}{\bar{\lambda}}\right) + \frac{1}{\ln(2)}\right]
\end{gather*}
Or, combining the sum
\begin{gather*}
     \frac{\partial I(S)}{\partial\lambda(s')} = \sum_{s \in S}\frac{p(s)p(s')}{\bar{\lambda}^2}\left((\delta_{s,s'}\bar{\lambda} - \lambda(s))\left[\log_2 \left(\frac{\lambda(s)}{\bar{\lambda}}\right) + \frac{1}{\ln(2)}\right]\right)
\end{gather*}
where $\delta_{s,s'}$ is the Kronecker delta, equal to 1 when $s = s'$ and zero otherwise.

\end{const}}

\subsection{Proof of Theorem \ref{thm:max_stiminfo}}
\label{pf:max_stiminfo}
\begin{proof}
    Let $\alpha(s) = \frac{\lambda_A(s)}{\Bar{\lambda}_A}$. Then, $\alpha(s) \ge 0$ and $\sum_{s \in S}\alpha(s)p(s) = 1$. Then, equation \ref{spatial information rate} becomes 
\begin{align*}
    I_{spike}(A: S) =  \sum_{s \in S} \alpha(s)p(s)\log_2 \left(\alpha(s)\right) = D_{KL}(q(s)\ ||\ p(s))
\end{align*}
where $q(s) = \alpha(s)p(s)$ and $D_{KL}$ is the Kullback-Leiber divergence. This grows when $q(s)$ becomes more concentrated compared to $p(s)$ and is maximized when $q(s)$ is completely concentrated to a single state $s^*$ [\cite{bonnici2020kullback}]. That is
\begin{align*}
    q(s) = \begin{cases}
        1 & s = s^*\\
        0 & s \neq s^*
    \end{cases}
\end{align*}
In this scenario, we have $\lambda_A(s^*)p(s^*) = \Bar{\lambda}_A$, thus,
\begin{align*}
    I_{spike}(A: S) &= \sum_{s \in S} \frac{\lambda_A(s)}{\Bar{\lambda}_A} \log_2 \left(\frac{\lambda_A(s)}{\Bar{\lambda}_A}\right) p(s) = \frac{\lambda_A(s^*)}{\lambda_A(s^*) p(s^*)}\log\left(\frac{\lambda_A(s^*)}{\lambda_A(s^*) p(s^*)}\right)p(s^*)\\
    &= \log_2\left(\frac{1}{p(s^*)}\right)
\end{align*}
This is clearly largest when the chosen state $s^* = \arg \min_{s \in S} p(s)$ and $p(s^*) = \epsilon$. 

\end{proof}

\subsection{Proposition \ref{prop:no-overlap} and Proof}\label{pf:no-overlap}

\begin{prop} Suppose for neurons $A$ and $B$ that $\{s\in S: \lambda_A(s) >0\} \cap \{s\in S:\lambda_B(s) > 0\} = \emptyset$. Then,
\begin{align*}
    &I_{sec}(A,B:S) = I_{sec}(A:S) + I_{sec}(B:S)\\
    &I_{spike}(A,B:S) = \frac{1}{\Lambda_{A,B}}(I_{sec}(A:S) + I_{sec}(B:S)).
\end{align*}
That is, the joint stimulus information rate depends only on the stimulus information rates of $A$ and $B$, respectively.
\label{prop:no-overlap}
\end{prop}

\begin{proof} If $\{s\in S: \lambda_A(s) >0\} \cap \{s\in S:\lambda_B(s) > 0\} = \emptyset$, then $\lambda_{A,B}(s) = \lambda_A(s)\lambda_B(s) = 0$ for all $s \in S$. Therefore, $\tilde{\lambda}_{A,B} = \sum_{s\in S}p(s)\sqrt{\lambda_{A,B}(s)} = 0$ and $\Bar{\lambda}_{A,B} = \sum_{s\in S}p(s)\lambda_{A,B}(s) = 0$. Thus, 
\begin{align*}
    I_{sec}(A,B: S) = &\sum_{s \in S} \Bigg[ p(s) 0
    + p(s)\left(\lambda_A(s) - 0\right)\log_2\left(\frac{\lambda_A(s) - 0}{\bar{\lambda}_A - 0}\right) + p(s)\left(\lambda_B(s) -0\right)\log_2\left(\frac{\lambda_B(s) -0}{\bar{\lambda}_B - 0}\right) \Bigg]
    \\ 
    = &\sum_{s \in S}p(s)\left(\lambda_A(s)\right)\log_2\left(\frac{\lambda_A(s)}{\bar{\lambda}_A}\right) + \sum_{s \in S}p(s)\left(\lambda_B(s)\right)\log_2\left(\frac{\lambda_B(s)}{\bar{\lambda}_B}\right)\\
    &= I_{sec}(A: S) + I_{sec}(B: S)
\end{align*}
And 
\begin{align*}
    I_{spike}(A,B:S) &= \frac{1}{\Lambda_{A,B}}I_{sec}(A,B: S)\\
    &=\frac{1}{\Lambda_{A,B}}(I_{sec}(A: S) + I_{sec}(B: S))
\end{align*}
    
\end{proof}

\subsection{Theorem \ref{thm:max_jointstiminfo} and Proof}
\label{pf:max_jointstiminfo}

\begin{theorem} For a fixed occupancy distribution with smallest entries $p(s_1) \ge p(s_2) > 0$, the maximum of the joint stimulus information rate is
\begin{align*}
    2(\bar{\lambda}_A + \bar{\lambda}_B)^{-1}\left[p(s_1)\lambda_A(s_1)\log_2(p(s_1)) + p(s_2)\lambda_B(s_2)\log_2(p(s_2))\right].
\end{align*}
In particular, for a uniform occupancy distribution, the maximum of the joint stimulus information rate is $2\log_2(n_s)$. Finally, this maximum is achieved precisely when neurons $A$ and $B$ fire in distinct stimulus bins. That is, the maximum is achieved when, without loss of generality, $\{s\in S: \lambda_A(s) > 0\} = s_i$ and $\{s\in S: \lambda_B(s) > 0\} = s_j$ for some $s_i, s_j \in S$ where $i \neq j$.
\label{thm:max_jointstiminfo}
\end{theorem}

\begin{proof} 
First, we may find the maximum of $I_{sec}(A, B:S)$. Now, recall
\begin{align*}
    I_{sec}(A,B) &= \sum_s p(s) r\sqrt{\lambda_{A,B}(s)}\log_2\left(\frac{\sqrt{\lambda_{A,B}(s)}}{\tilde{\lambda}_{A,B}} \right)\\ 
    &+ p(s)\left(\lambda_A(s) - r\sqrt{\lambda_{A,B}(s)}\right)\log_2\left(\frac{\lambda_A(s) - r\sqrt{\lambda_{A,B}(s)}}{\bar{\lambda}_A - r\tilde{\lambda}_{A,B}}\right)\\
    &+ p(s)\left(\lambda_B(s) - r\sqrt{\lambda_{A,B}(s)}\right)\log_2\left(\frac{\lambda_B(s) - r\sqrt{\lambda_{A,B}(s)}}{\bar{\lambda}_B - r\tilde{\lambda}_{A,B}}\right)
\end{align*}
where $\sum_s p(s) \sqrt{\lambda_{A,B}}(s) = \tilde{\lambda}_{A,B}$, and $\lambda_A(s)\lambda_B(s) = \lambda_{A,B}(s)$. The second and third terms dominate this sum, and are maximized when 
\begin{align*}
    \lambda_A(s_1) - r\sqrt{\lambda_{A,B}(s_1)} \neq 0\\ 
    \lambda_B(s_2) - r\sqrt{\lambda_{A,B}(s_2)} \neq 0 
\end{align*}
for particular $s_1, s_2 \in S$ are zero for all remaining $s\in S$. That is, for all remaining $s\in S$
\begin{align*}
    \lambda_A(s) = \lambda_B(s) =  r\sqrt{\lambda_{A,B}(s)}
\end{align*}
If $s_1 = s_2$, then $r = 1$ and $\epsilon_1 = \epsilon_2$ and 
\begin{align*}
    I_{sec}(A,B:S) &=  \epsilon_1\lambda_A(s_1)\log(1/\epsilon_1) + \epsilon_1\lambda_B(s_1)\log_2(1/\epsilon_1)- \epsilon_1\sqrt{\lambda_A(s_1)\lambda_B(s_1)}\log_2(1/\epsilon_1) \\
    &\leq \epsilon_1\lambda_A(s_1)\log(1/\epsilon_1) + \epsilon_1\lambda_B(s_1)\log_2(1/\epsilon_1)
\end{align*}
and 
\begin{align*}
    I_{spike}(A,B:S) & \leq 2\cdot\frac{\epsilon_1\lambda_A(s_1)\log(1/\epsilon_1) + \epsilon_1\lambda_B(s_1)\log_2(1/\epsilon_1)}{\Bar{\lambda}_A + \Bar{\lambda}_B}  = 2\log_2(1/\epsilon_1)
\end{align*}
If $s_1 \neq s_2$ then $r = -1$ and $\lambda_{A,B}(s) = 0$ for all $s\in S$. In this case, 
\begin{align*}
    I_{sec}(A,B:S) &= \sum_s p(s)\lambda_A(s)\log_2\left(\frac{\lambda_A(s)}{\bar{\lambda}_A}\right) +  p(s)\lambda_B(s)\log_2\left(\frac{\lambda_B(s)}{\bar{\lambda}_B }\right)\\
    &= I_{sec}(A:S) + I_{sec}(B:S)
\end{align*}
By Theorem \ref{thm:max_stiminfo}, the resulting maximum is 
\begin{align*}
    \epsilon_1\lambda_A(s_1)\log(1/\epsilon_1) + \epsilon_2\lambda_B(s_2)\log_2(1/\epsilon_2)
\end{align*} 
with $\epsilon_1\lambda_A(s_1) = \Bar{\lambda}_A$ and $\epsilon_2\lambda_B(s_2) = \Bar{\lambda}_B$. Thus, for $\epsilon_2 \in (0,1)$ with $\epsilon_2 \ge \epsilon_1$, 
\begin{align*}
    I_{spike}(A,B:S) &= 2\cdot\frac{\epsilon_1\lambda_A(s_1)\log(1/\epsilon_1) + \epsilon_2\lambda_B(s_2)\log_2(1/\epsilon_2)}{\Bar{\lambda}_A + \Bar{\lambda}_B}\\
    &= 2\cdot\frac{\Bar{\lambda}_A\log(1/\epsilon_1) + \Bar{\lambda}_B\log_2(1/\epsilon_2)}{\Bar{\lambda}_A + \Bar{\lambda}_B}
    \ge 2\cdot\log(1/\epsilon_1)
\end{align*} 
Thus, the maximum is achieved when $s_1 \neq s_2$, i.e. neurons fire in single and separate stimulus bins. The resulting joint stimunlus information rate, is
\begin{align*}
    I_{spike}(A,B:S)  =\frac{2\left(\epsilon_1\lambda_A(s_1)\log_2(1/\epsilon_1) + \epsilon_2\lambda_B(s_2)\log_2(1/\epsilon_2)\right)}{\Bar{\lambda}_A + \Bar{\lambda}_B}
\end{align*}
\end{proof}

\subsection{Proposition \ref{prop:no_intersect} and Proof}
\label{pf:no_intersect}

\begin{prop} Suppose we have a stimulus space with $n_s$ stimulus bins and a uniform occupancy distribution. If $\{s\in S: \lambda_{A_i}(s) >0\} \cap \{s\in S:\lambda_{A_j}(s) > 0\} = \emptyset$ for all $i,j = 1,\ldots, n_p,\ i\neq j$, then
\begin{align*}
    J = \sum_{i  = 1}^{n_p} I_{sec}(A_i)(e_iv_i^T + v_ie_i^T),
\end{align*}
where $v_i = 2\left[\frac{1}{2(\bar{\lambda}_{A_1} + \bar{\lambda}_{A_i})}, \frac{1}{\bar{\lambda}_{A_2} + \bar{\lambda}_{A_i}}, \cdots \frac{1}{\bar{\lambda}_{A_i} + \bar{\lambda}_{A_i}}, \cdots \frac{1}{\bar{\lambda}_{A_{n_p}} + \bar{\lambda}_{A_i}}\right]^T$.
    \label{prop:no_intersect}
\end{prop}

\begin{proof} Using Proposition \ref{prop:no-overlap},
\begin{align*}
    J_{i,j} = I_{spike}(A_i,A_j:S) =\frac{1}{\Lambda_{A_i,A_j}}(I_{sec}(A_i: S) + I_{sec}(A_j: S))
\end{align*}
We may decompose $J$ into a sum of $n_p$ matrices, where each matrix contains only containing a factor of $I_{sec}(A_i)$ for $i = 1,\ldots, n_p$ and is all zero everywhere else. 
\begin{align*}
    &J = I_{sec}(A_1)(e_1v_1^T + v_1e_1^T) + \ldots + I_{sec}(A_{n_p})(e_{n_p}v_{n_p}^T + v_{n_p}e_{n_p}^T)\\
    &J = \sum_{i  = 1}^{n_p} I_{sec}(A_i)(e_iv_i^T + v_ie_i^T)
\end{align*}
where 
\begin{align*}
    v_i = 2\left[\frac{1}{2(\Bar{\lambda}_{A_1} + \Bar{\lambda}_{A_i})}, \frac{1}{\Bar{\lambda}_{A_2} + \Bar{\lambda}_{A_i}}, \cdots \frac{1}{\Bar{\lambda}_{A_i} + \Bar{\lambda}_{A_i}}, \cdots \frac{1}{\Bar{\lambda}_{A_{n_p}} + \Bar{\lambda}_{A_i}}\right]^T
\end{align*}
 
\end{proof}

\subsection{Theorem \ref{thm:special_case} and Proof}
\label{pf:special_case}

\begin{theorem} Suppose we have a stimulus space with $n_s$ stimulus bins and a uniform occupancy distribution. Given $n_p = n_s$ neurons such that $\{s\in S: \lambda_{A_i}(s) > 0\} \neq \emptyset$ for all $i = 1,\ldots, n_p$ and $\{s\in S: \lambda_{A_i}(s) >0\} \cap \{s\in S: \lambda_{A_j}(s) > 0\} = \emptyset$ for all $i,j = 1,\ldots, n_p,\ i\neq j$, we have 
\begin{align*}
    J_{i,j} = \log_2(n_s)(2 - \delta_{i,j}),
\end{align*}
where $n_s$ is the number of stimulus bins and $\delta_{i,j}$ is the Kronecker delta, equal to 1 when $i = j$ and zero otherwise.
\label{thm:special_case}
\end{theorem}

\begin{proof}
There are two cases. 

First when $i = j$, by we have $I_{spike}(A_i, A_j: S) = I_{spike}(A_i, A_i:S) = I_{spike}(P_i:S)$. Since each neuron only fires in one stimulus bin, by Theorem \ref{thm:max_stiminfo}, $J_{i, i}= I_{spike}(A_i:S) = \log_2(n_s)$ for all $i = 1,\ldots, n_p$. 

Secondly, when $i \neq j$, we may invoke Proposition \ref{prop:no-overlap} together. That is, 
    \begin{align*}
        J_{i,j} &= \frac{2}{\Bar{\lambda}_{A_i} + \Bar{\lambda}_{A_j}}\left(\sum_{s \in S} \lambda_{A_i}(s) \log_2\left(\frac{\lambda_{A_i}(s)}{\Bar{\lambda}_{A_i}}\right)\frac{1}{n_s} + \lambda_{A_j}(s) \log_2\left(\frac{\lambda_{A_j}(s)}{\Bar{\lambda}_{A_j}}\right)\frac{1}{n_s} \right)\\
        &= \frac{2}{\Bar{\lambda}_{A_i} + \Bar{\lambda}_{A_j}}\left(\Bar{\lambda}_{A_i}\log_2(n_s) + \Bar{\lambda}_{A_j}\log_2(n_s)\right) = 2\log_2(n_s)
    \end{align*} 
    Together, we may write 
    \begin{align*}
        J_{i,j} = \log_2(n_s)(2 - \delta_{i,j})
    \end{align*}
    where $\delta_{i,j}$ is the Kronecker delta, equal to 1 when $i = j$ and zero otherwise. 
    
\end{proof}

\subsection{Corollary \ref{cor:eigenvalue_eigenvector} and Proof}
\label{pf:eigenvalue_eigenvector}

\begin{cor} Under the same assumptions as Theorem \ref{thm:special_case}, the leading eigenvalue and eigenvector pair  $(\lambda_1, v_1)$ of $J$ is 
\begin{align*}
    (\lambda_1, v_1) = (\log_2(n_s)(2n_p - 1),\ \textbf{1}),
\end{align*}
 where $\textbf{1} \in \R^{n_p}$ is the vector of all ones.
\label{cor:eigenvalue_eigenvector}
\end{cor}

\begin{proof}
    We can write 
    \begin{align*}
        J = \log_2(n_s)\begin{bmatrix} 
        1 & 2 & 2 & \dots & 2\\
        2 & 1 & 2 & \dots & 2\\
        2 & 2 & 1 & \dots & 2\\
        \vdots & \vdots & \vdots & \ddots & \vdots\\
        2 & 2 & 2 & \dots & 1
        \end{bmatrix} = \log_2(n_s)(2\textbf{1}\textbf{1}^T - I)
    \end{align*}
Now, since $I$ and $\textbf{1}\textbf{1}^T$ are diagonalizable and commute, they share the same eigenvectors. The only eigenvalue of $I$ is 1. The eigenvalues of $\textbf{1}\textbf{1}^T$ are $n_p$ with multiplicity $1$ and $0$ with multiplicity $n_p - 1$. Thus, the leading eigenvalue of $J$ is $\log_2(n_s)(2n_p - 1)$. 

The leading eigenvector corresponds to the eigenvector with eigenvalue $n_p$ of the unit matrix $\textbf{1}\textbf{1}^T$, which is the vector $\textbf{1}$. 
\end{proof}

\subsection{Proof of Theorem \ref{thm:largest_eigenvalue}}
\label{pf:largest_eigenvalue}

\begin{proof} Let $\Hat{J}$ be the matrix described in Theorem \ref{thm:special_case}. For a general stimulus information matrix $J$, by Theorem \ref{thm:max_jointstiminfo}, $J_{i,j} \leq \Hat{J}_{i,j}$ for all $i,j$. Now, since $J$ is symmetric with nonnegative entries, the leading eigenvector of $J$, $v$, has all positive entries and maximizes the Rayleigh Quotient of $J$. Thus, 
\begin{align*}
     \max_{x,\ ||x|| = 1} x^TJx = \lambda_1 = v^TJv < v^T\Hat{J}v \leq \max_{x,\ ||x|| = 1} x^T\Hat{J}x = \log_2(n_s)(2n_p - 1)
\end{align*}
So, the leading eigenvalue $J$ is less than that of $\Hat{J}$.
\end{proof}

\begin{figure}[!htb]
    \centering
    \includegraphics[width = 0.9\linewidth]{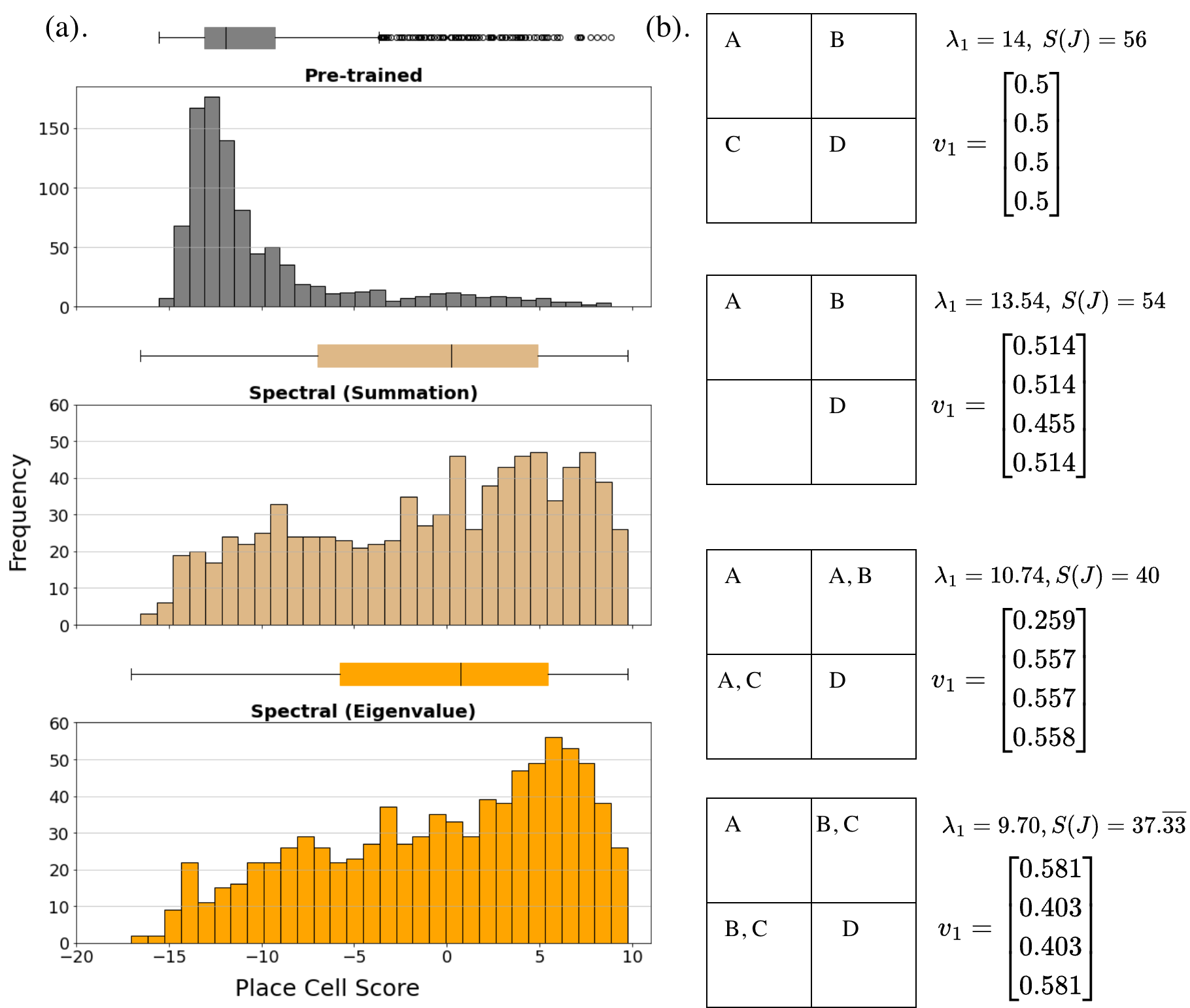}
    \caption{(a) Place cell box plots and histograms across 20 models (10 trained with 32 place cells, 10 with 64). Comparison in place cell score (Equation \eqref{place cell score}) is shown between models trained via maximizing $S(J)$, the summation of the entries of $J$ versus maximizing the leading eigenvalue of $J$. (b) Comparison between the leading eigenvalue of the spatial information matrix $J$, $\lambda_1$, and $S(J)$ across four scenarios. Consider four neurons, A, B, C, and D in a simple four-quadrant square arena. If the letter of a neuron appears in a quadrant, then we set the firing rate of the neuron in that quadrant to $1$ spike/sec. The corresponding leading eigenpair $(\lambda_1, v_1)$ as well as $S(J)$ are shown for each scenario.}
    \label{fig:summation_vs_eigenvalue}
\end{figure}

\begin{rem}[Eigenvalues of Spatial Information Matrix]
\label{evals}

Suppose we have a data matrix $P \in \R^{n_x \times n_p}$ representing $n_p$ place cell firing rates across a trajectory of length $n_x$. That is, $P_{i,j} = \lambda_{j}(x_i)$, place cell $j$'s mean firing rate at spatial location $x_i$. From $P$, we construct $J \in \R^{n_p \times n_p}$ with $J_{i,j} = I_{spike}(P_i, P_j)$, the joint spatial information rate between place cells $i,j,$ and the trajectory. Let, $R(J,v)$ be the the Rayleigh quotient [\cite{horn1990matrix}] of $J$ with nonzero vector $v$ such that $||v||_2 = 1$, and $S(J)$ be the sum of the entries of $J$. That is, 
\begin{align*}
    R(J,v) = v^TJv &= \sum_{i = 1}^{n_p} \sum_{j = 1}^{n_p} v_i J_{ij} v_j\\
    S(J) &=  \sum_{i = 1}^{n_p} \sum_{j = 1}^{n_p} J_{ij}
\end{align*}
It follows that $\underset{v,\ ||v||_2 = 1}{\text{max}} R(J,v) = \lambda_1$ where $\lambda_1$ is the leading eigenvalue of $J$. When $v_i > 0 $ for all $i = 1,\ldots, n_p$, we have that $\lambda_1$ is positively correlated with $S(J)$ (Figure \ref{fig:summation_vs_eigenvalue}).

\end{rem}

\subsection{Partial Information Decomposition}
\label{PID_supplemetnary}

The \textbf{redundancy} between two neurons $A,B$ is 
\begin{align*}
    \text{Rdn}(A,B:S) = \sum_s p(s) \min_{x \in \{A,B\}} I_{spec}(x:s),
\end{align*}
where $I_{spec}$ is the specific information \cite{deweese1999measure}, i.e.\ the amount of information provided by a neuron about a specific state $s \in S$ of the stimulus. This is given by
\begin{align*}
    I_{spec}(x:s) = \frac{\lambda_x(s)}{\Bar{\lambda}_x} \log_2 \left(\frac{\lambda_x(s)}{\Bar{\lambda}_x}\right). 
\end{align*}
Secondly, the \textbf{unique information} of a neuron $A$ with respect to neuron $B$ is 
\begin{align*}
    \text{Unq}(A:S) = I_{spike}(A:S) - \text{Rdn}(A,B:S).
\end{align*}
Now, using our novel joint stimulus information rate, we define the \textbf{synergy} between two neurons $A,B$ as 
\begin{align*}
    \text{Syn}(A,B:S) = I_{spike}(A,B:S) - I_{spike}(A:S) - I_{spike}(B:S) + \text{Rdn}(A,B:S),
\end{align*}
and the \textbf{redundancy-synergy index} between two neurons $A, B$ as
\begin{align*}
    \text{RS}(A,B:S) &= \text{Syn}(A,B:S) - \text{Rdn}(A,B:S)\\
    &= I_{spike}(A,B:S) - I_{spike}(A:S) - I_{spike}(B:S).
\end{align*}

%%%%%%%%%%%%%%%%%%%%%%%%%%%%%%%%%%%%%%%%%%%%%%
%% Supplementary Material, including data   %%
%% sets and code, should be provided in     %%
%% {supplement} environment with title      %%
%% and short description. It cannot be      %%
%% available exclusively as external link.  %%
%% All Supplementary Material must be       %%
%% available to the reader on Project       %%
%% Euclid with the published article.       %%
%%%%%%%%%%%%%%%%%%%%%%%%%%%%%%%%%%%%%%%%%%%%%%
\newpage
\vspace*{\fill}

\begin{figure}[!htb]
    \centering
    \includegraphics[width = \linewidth]{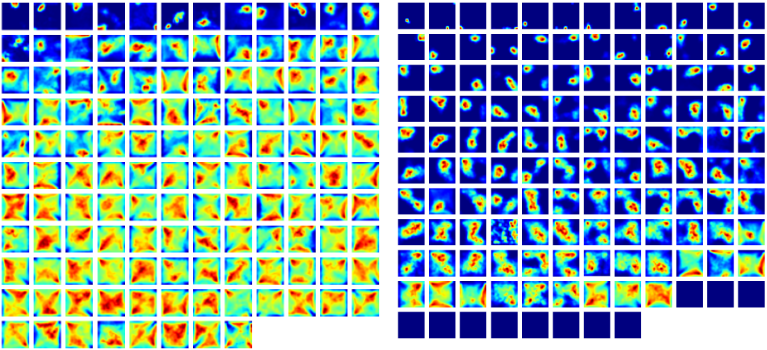}
    \caption{Untrained and trained (with spectral-spatial information) place cell activations for 128 place cells. Each place cell's activations are normalized between 0 and 1. The top 64 are displayed in Figure \ref{fig:top64}. }
    \label{fig:all_activations}
\end{figure}
\vspace*{\fill}
\newpage

\newpage
\vspace*{\fill}

%\section{Joint Spatial Information Rate on real place cell recordings}

\begin{figure}[!htb]
    \centering
    \includegraphics[width = 0.9\textwidth]{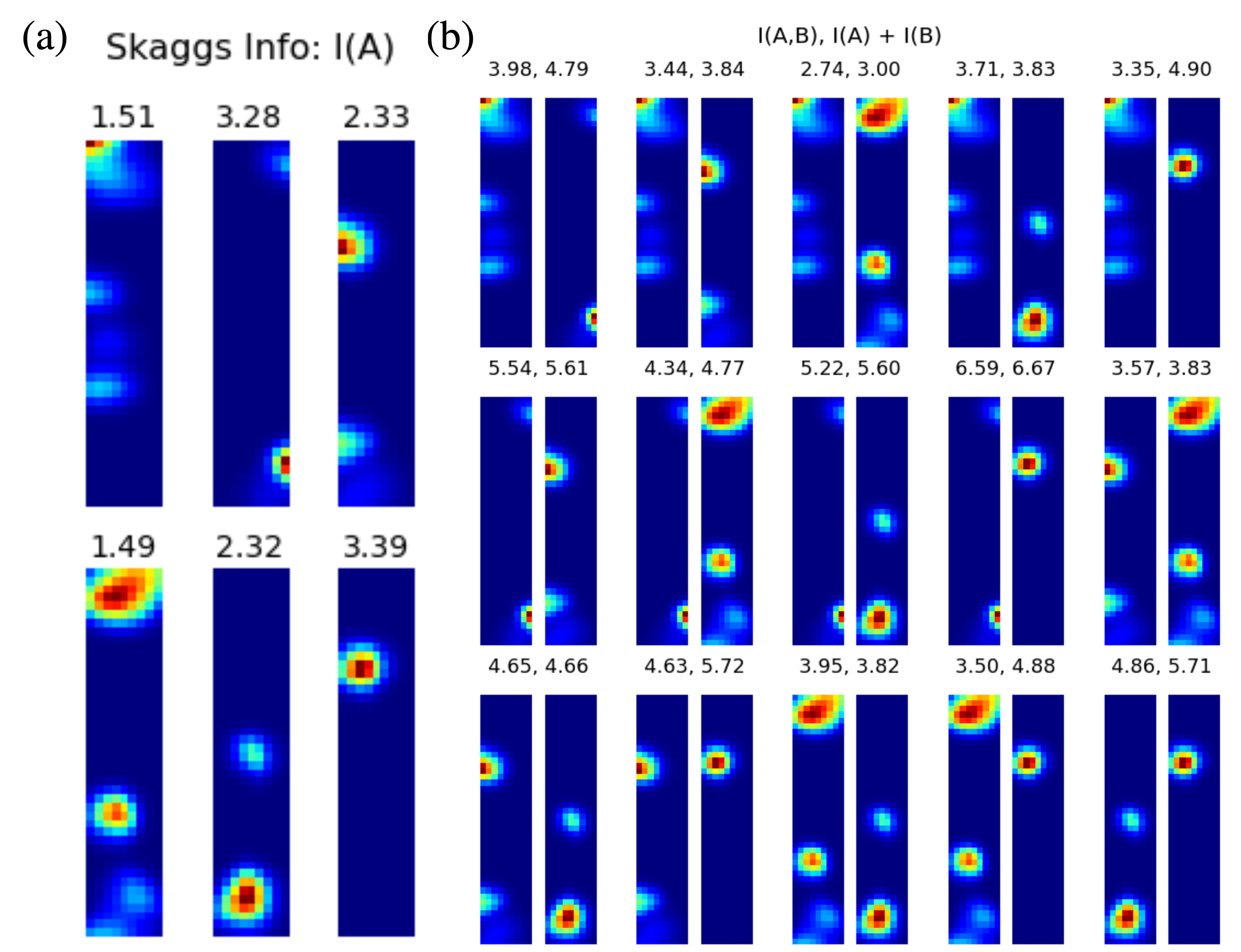}
    \caption{(a) Skaggs' spatial information rate for six neurons from \cite{hazama2019data}. (b) Joint spatial information, $I(A, B)$, compared to the sum of Skaggs' spatial information, $I(A) + I(B)$, for all possible pairs of neurons. The leading eigenvalue of the corresponding spatial information matrix is $\lambda_1 = 24.21$.}
    \label{fig:JSIR_true_recording}
\end{figure}
\vspace*{\fill}
\newpage
%\section{Joint Spatial Information Rate Non-negativity}

\newpage
\vspace*{\fill}
\begin{figure}[!h]
    \centering
    \includegraphics[width=\textwidth]{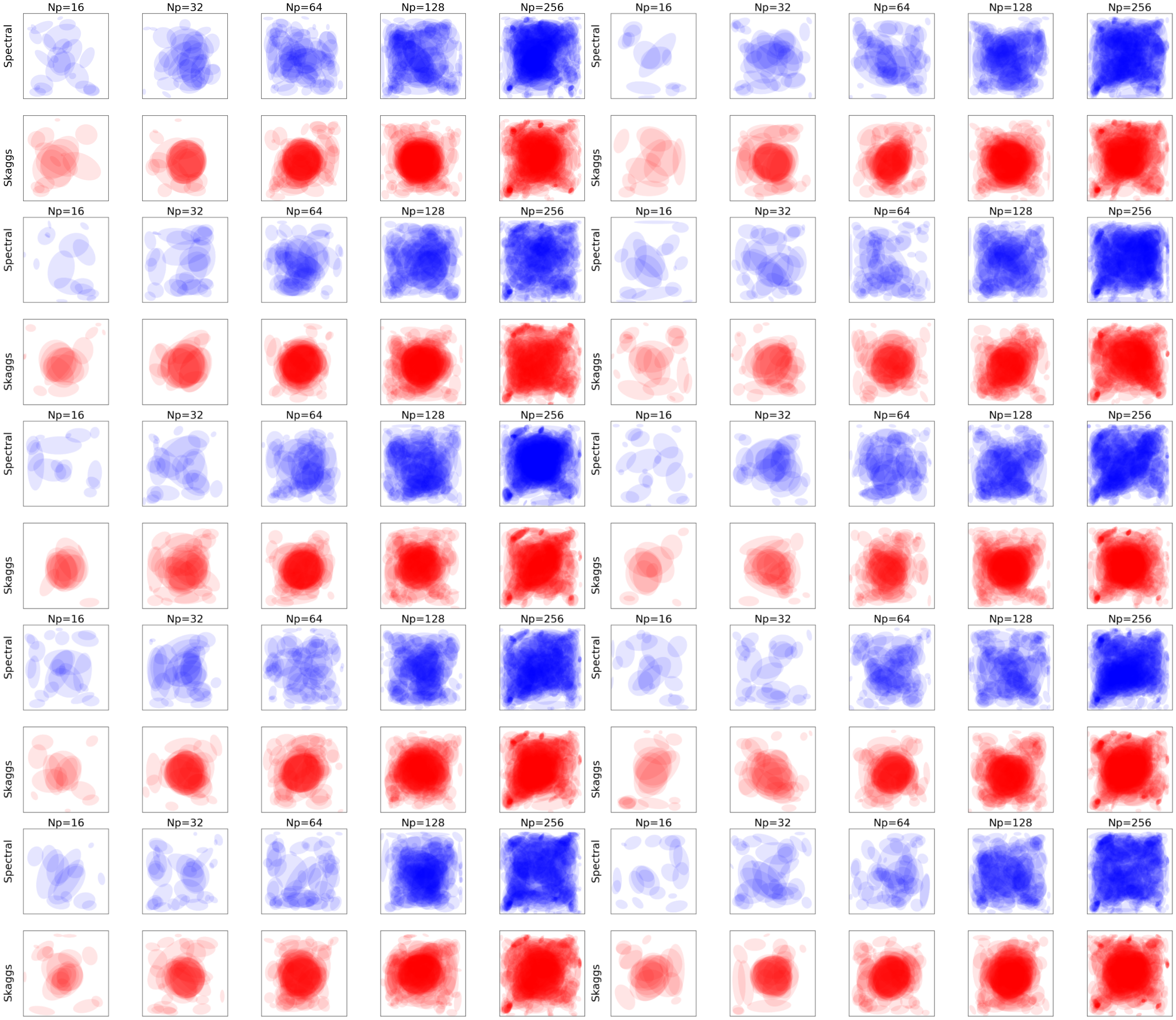}
    \caption{A comparison of the distribution of place fields between models trained with spectral-spatial information versus Skaggs information across 10 experiments.}
    \label{fig:heatmaps_all}
\end{figure}

\vspace*{\fill}
\newpage

\newpage
\vspace*{\fill}
\begin{figure}[!h]
    \centering
    \includegraphics[width=\textwidth]{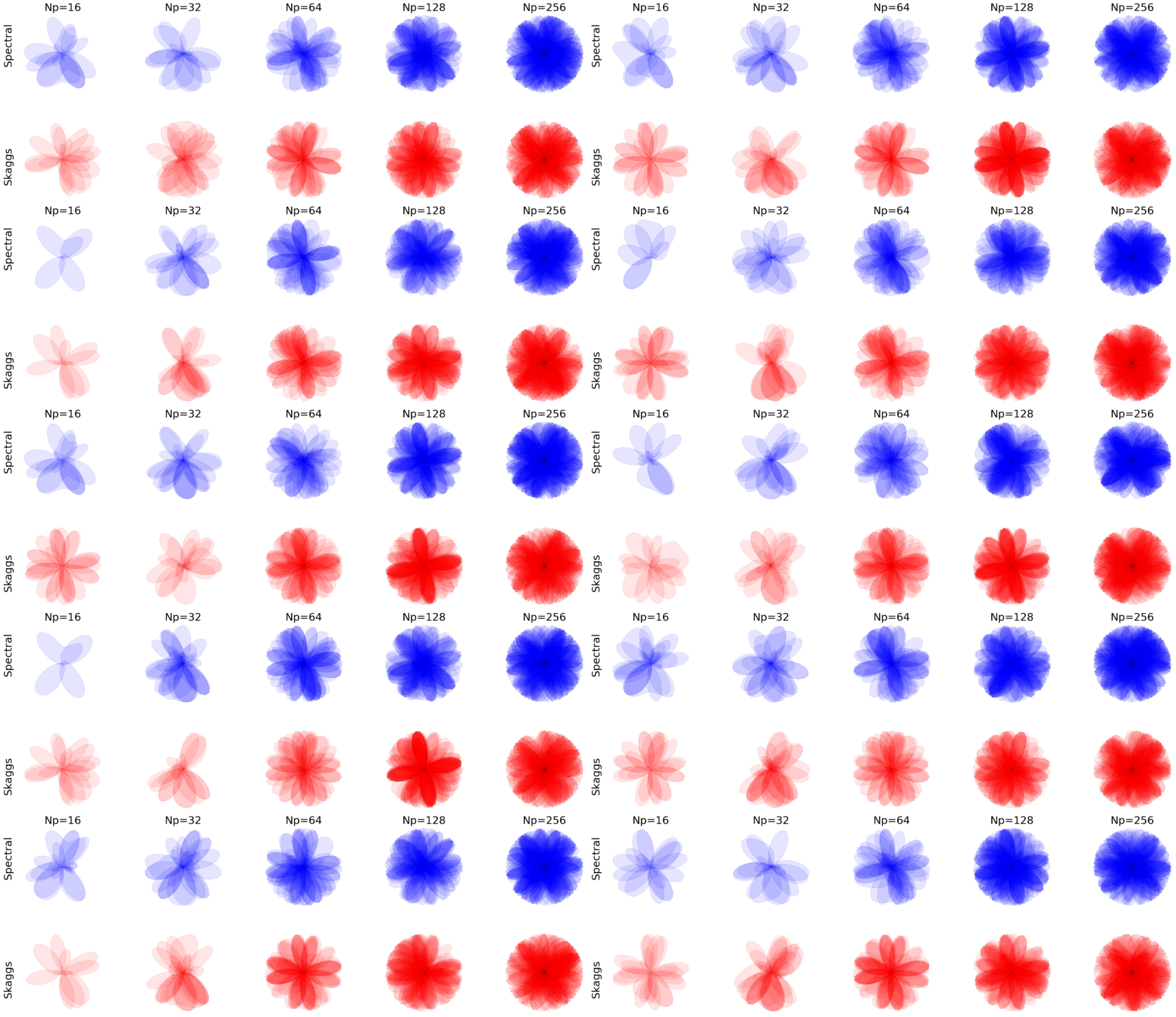}
    \caption{A comparison of the distribution of preferred firing angles between models trained with spectral-directional information versus Skaggs information across 10 experiments.}
    \label{fig:hd_heatmaps_all}
\end{figure}

\vspace*{\fill}
\newpage

\newpage
\vspace*{\fill}
\begin{figure}[!h]
    \centering
    \includegraphics[width = 0.4\textwidth]{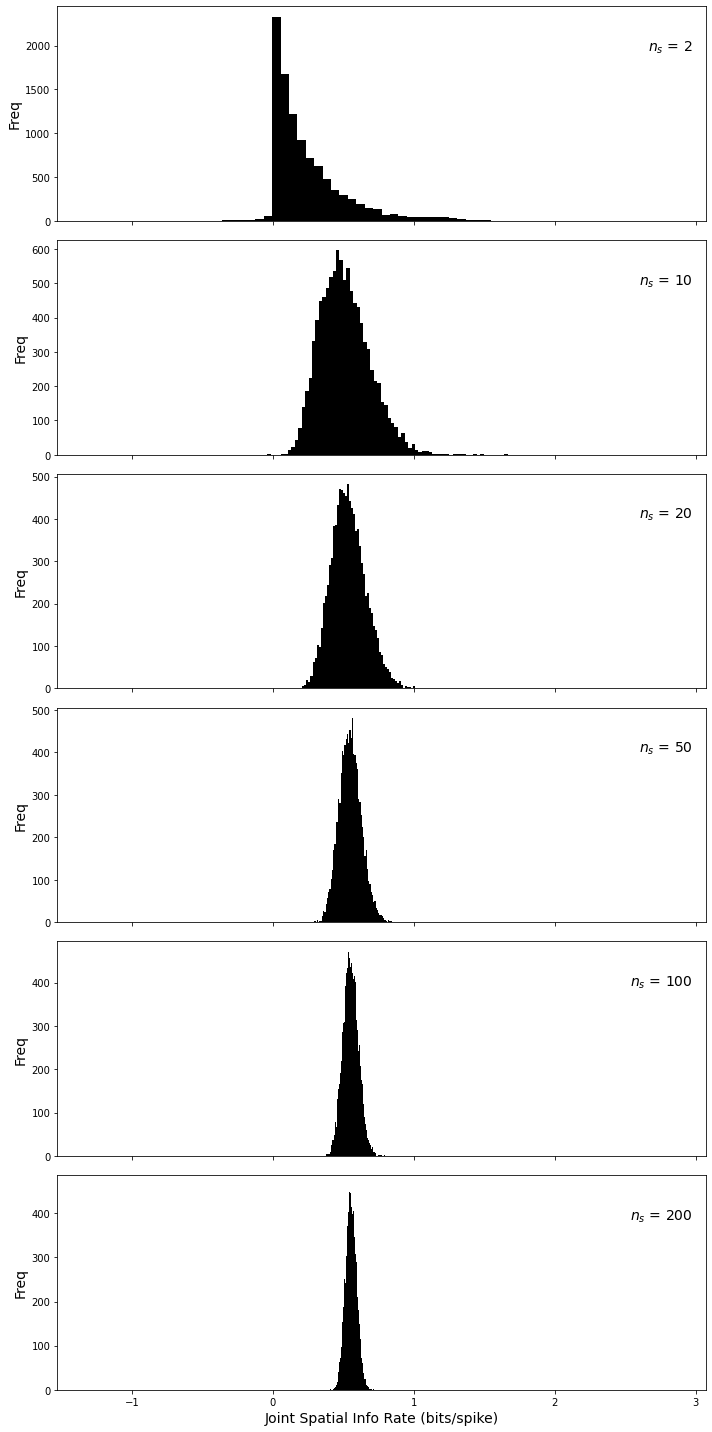}
    \caption{Histograms of the joint spatial information rate (bits/spike) from 1000 pairs of neurons for $n_s = 2, 10, 20, 50, 100,$ and $200$. For each neuron, firing rates are randomly drawn from a continuous uniform distribution with a minimum of 0 and a maximum of 200. Our experiments suggest that the joint spatial information rate is nonnegative for $n_s \geq 20$.}
    \label{fig:info_hists}
\end{figure}

\begin{figure}[!h]
    \centering
    \includegraphics[width =\textwidth]{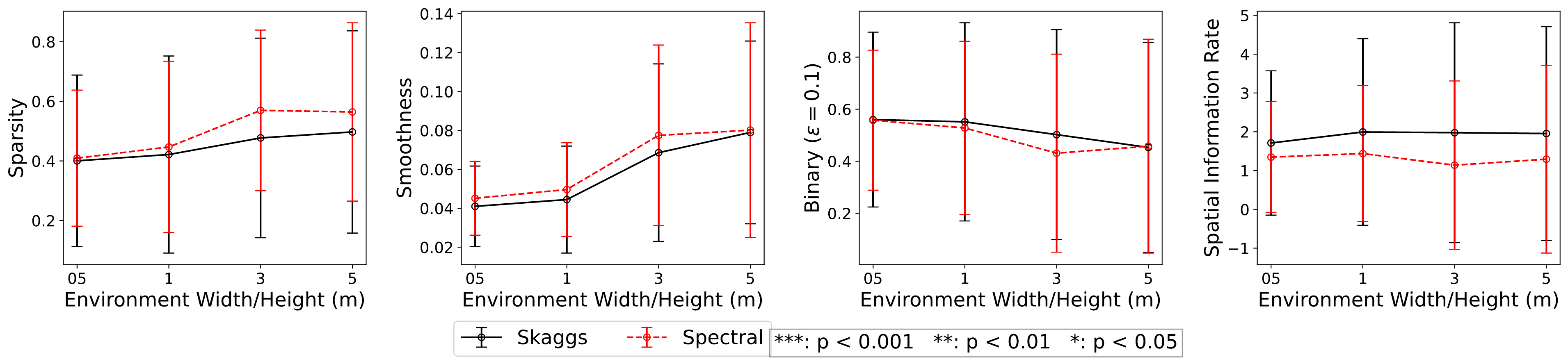}
    \caption{Mean ($\pm$ standard deviation) sparsity, smoothness, binarity ($\epsilon = 0.1$), and spatial information rate of 256 trained place cells as the environment size increases.}
    \label{fig:env_metrics_plot}
\end{figure}

\newpage
\vspace*{\fill}

\begin{figure}[!h]
    \centering
    \includegraphics[width =\textwidth]{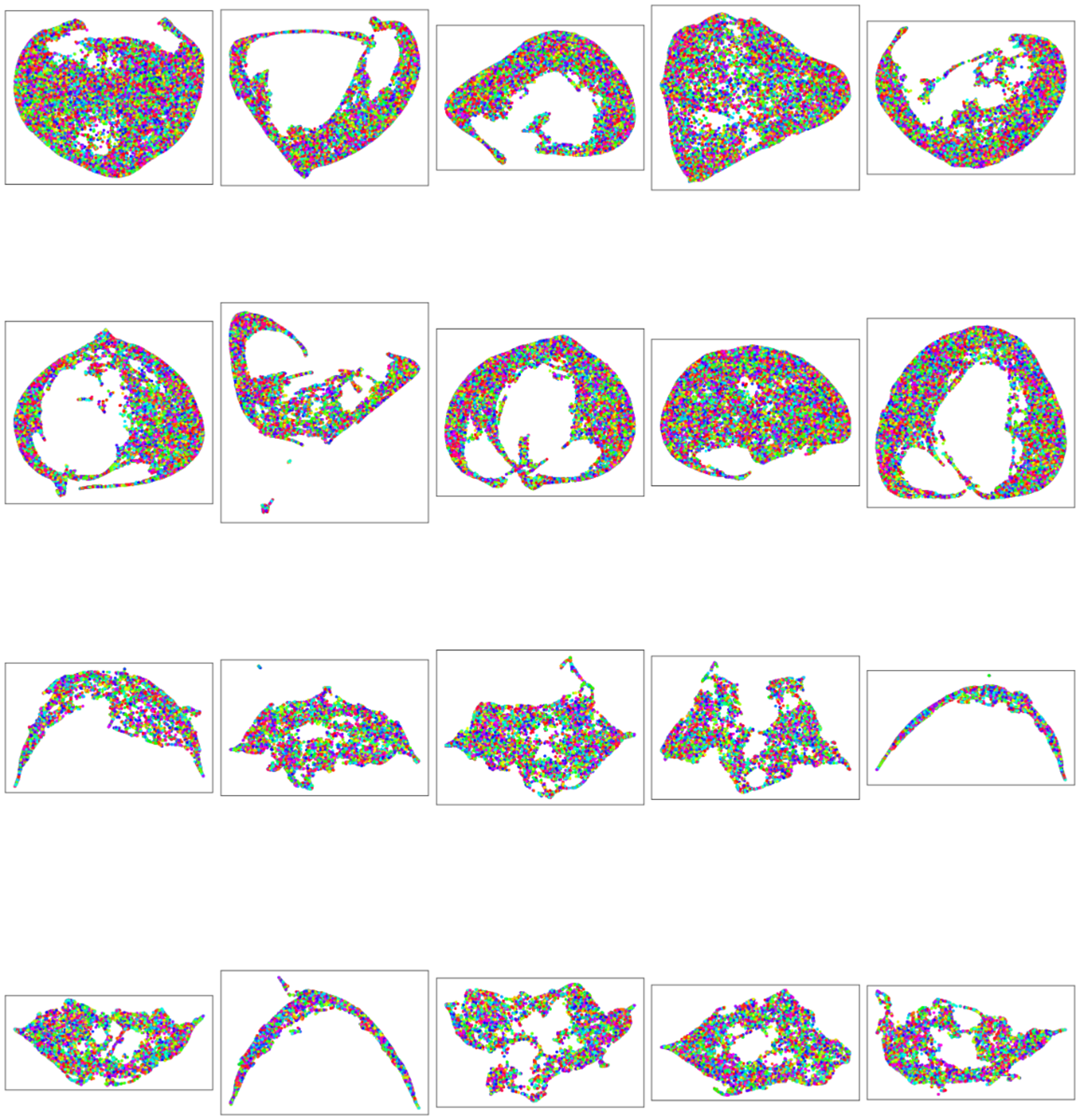}
    \caption{Results of dimension reduction applied to learned head direction cell activations form spectral-trained models with 64 and 128 neurons. First, UMAP [\cite{mcinnes2018umap}] was applied to reduce dimensionality to 10, followed by principle component analysis to 2 dimensions.}
    \label{fig:rings}
\end{figure}
\vspace*{\fill}
\newpage

\section{Sensitivity Analysis}
\label{app:sensitivity}
To test the effect of training with spectral-spatial information as compared to Skaggs spatial information rate across training parameters, we train three distinct models (each with 32 place cells) for all possible combinations of the following parameters:
\begin{itemize}
    \item Arena Size (in meters): 0.25, 0.75, 1, 1.5
    \item Number of Hidden Units (Ng): 256, 512, 1028
    \item Learning Rate: 0.001, 0.0001, 0.00001
    \item Batch Size: 10, 40, 80
    \item Sequence Length: 40, 80, 120
\end{itemize}
We implement a simple stopping criterion during training, in which training halts if the current epoch loss is greater than the mean of the previous three epoch losses (provided the model has trained for at least five epochs). The mean place cell score ($\pm$ standard deviation) is presented below across pairs of parameters by averaging over remaining parameters.

\begin{minipage}{0.48\textwidth}
\centering
    \textbf{Skaggs PC Score} ($N_p  = 32$)
\end{minipage}
\hfill
\begin{minipage}{0.48\textwidth}
\centering
    \textbf{Spectral PC Score} ($N_p  = 32$)
\end{minipage}
\hfill
\begin{minipage}{0.48\textwidth}
\centering
\resizebox{\textwidth}{!}{\begin{tabular}{ccccc}
\toprule
Arena Size (m) & 0.25 & 0.75 & 1 & 1.5 \\
Ng &  &  &  &  \\
\midrule
256 & -2.81 ± 6.01 & -4.22 ± 6.04 & -3.24 ± 5.98 & -1.52 ± 4.61 \\
512 & -1.93 ± 5.45 & -3.65 ± 5.36 & -2.6 ± 5.52 & -0.48 ± 3.99 \\
1028 & -1.67 ± 5.22 & -2.98 ± 5.15 & -2.11 ± 5.5 & 0.42 ± 3.64 \\
\bottomrule
\end{tabular}
}
\end{minipage}
\hfill
\begin{minipage}{0.48\textwidth}
\centering
\resizebox{\textwidth}{!}{\begin{tabular}{c|cccc}
\toprule
Arena Size (m) & 0.25 & 0.75 & 1 & 1.5 \\
Ng &  &  &  &  \\
\midrule
256 & \textbf{-1.6 ± 5.54} & \textbf{-2.9 ± 6.25} & \textbf{-2.21 ± 5.93} & \textbf{-0.61 ± 4.47} \\
512 & \textbf{-0.46 ± 5.48} & \textbf{-2.27 ± 5.96} & \textbf{-1.22 ± 5.91} & \textbf{0.2 ± 4.1} \\
1028 & \textbf{-0.33 ± 5.52} & \textbf{-2.35 ± 5.3} & \textbf{-1.28 ± 5.76} & \textbf{0.68 ± 3.96} \\
\bottomrule
\end{tabular}
}
\end{minipage}

\begin{minipage}{0.48\textwidth}
\centering
\resizebox{\textwidth}{!}{\begin{tabular}{ccccc}
\toprule
Learning Rate & 1e-03 & 1e-04 & 1e-05 \\
Ng &  &  &  \\
\midrule
256 & \textbf{-6.62 ± 5.49} & -0.16 ± 5.45 & -2.07 ± 6.03 \\
512 & \textbf{-4.48 ± 5.26} & -1.43 ± 4.92 & -0.58 ± 5.06 \\
1028 & \textbf{-1.93 ± 4.96} & -1.65 ± 4.93 & -1.17 ± 4.75 \\
\bottomrule
\end{tabular}
}
\end{minipage}
\hfill
\begin{minipage}{0.48\textwidth}
\centering
\resizebox{\textwidth}{!}{\begin{tabular}{c|cccc}
\toprule
Learning Rate & 1e-03 & 1e-04 & 1e-05 \\
Ng &  &  &  \\
\midrule
256 & -6.79 ± 5.3 & \textbf{2.16 ± 5.75} & \textbf{-0.86 ± 5.59} \\
512 & -4.78 ± 5.17 & \textbf{1.03 ± 5.96} & \textbf{0.93 ± 4.96} \\
1028 & -2.25 ± 4.76 & \textbf{-1.28 ± 5.46} & \textbf{1.09 ± 5.18} \\
\bottomrule
\end{tabular}
}
\end{minipage}

\begin{minipage}{0.48\textwidth}
\centering
\resizebox{\textwidth}{!}{\begin{tabular}{ccccc}
\toprule
Batch Size & 10 & 40 & 80 \\
Ng &  &  &  \\
\midrule
256 & -6.3 ± 6.66 & -4.54 ± 6.01 & 1.99 ± 4.31 \\
512 & -6.09 ± 6.32 & -3.75 ± 5.59 & 3.34 ± 3.34 \\
1028 & -5.68 ± 6.38 & -3.16 ± 5.47 & 4.08 ± 2.79 \\
\bottomrule
\end{tabular}
}
\end{minipage}
\hfill
\begin{minipage}{0.48\textwidth}
\centering
\resizebox{\textwidth}{!}{\begin{tabular}{c|cccc}
\toprule
Batch Size & 10 & 40 & 80 \\
Ng &  &  &  \\
\midrule
256 & \textbf{-4.76 ± 6.37} & \textbf{-3.16 ± 6.08} & \textbf{2.43 ± 4.19} \\
512 & \textbf{-4.1 ± 6.67} & \textbf{-2.42 ± 5.95} & \textbf{3.71 ± 3.47} \\
1028 & \textbf{-4.48 ± 6.55} & \textbf{-2.16 ± 5.8} & \textbf{4.2 ± 3.05} \\
\bottomrule
\end{tabular}
}
\end{minipage}

\begin{minipage}{0.48\textwidth}
\centering
\resizebox{\textwidth}{!}{\begin{tabular}{ccccc}
\toprule
Sequence Length & 40 & 80 & 120 \\
Ng &  &  &  \\
\midrule
256 & -1.87 ± 5.28 & -2.76 ± 5.68 & -4.22 ± 6.02 \\
512 & -1.25 ± 4.87 & -2.1 ± 4.95 & -3.15 ± 5.42 \\
1028 & -1.03 ± 4.81 & -1.52 ± 4.84 & -2.21 ± 4.99 \\
\bottomrule
\end{tabular}
}
\end{minipage}
\hfill
\begin{minipage}{0.48\textwidth}
\centering
\resizebox{\textwidth}{!}{\begin{tabular}{c|cccc}
\toprule
Sequence Length & 40 & 80 & 120 \\
Ng &  &  &  \\
\midrule
256 & \textbf{-0.82 ± 5.26} & \textbf{-1.94 ± 5.65} & \textbf{-2.73 ± 5.73} \\
512 & \textbf{-0.22 ± 5.13} & \textbf{-0.88 ± 5.31} & \textbf{-1.7 ± 5.65} \\
1028 & \textbf{-0.43 ± 5.0} & \textbf{-0.82 ± 5.14} & \textbf{-1.2 ± 5.27} \\
\bottomrule
\end{tabular}
}
\end{minipage}

\begin{minipage}{0.48\textwidth}
\centering
\resizebox{\textwidth}{!}{\begin{tabular}{ccccc}
\toprule
Learning Rate & 1e-03 & 1e-04 & 1e-05 \\
Arena Size (m) &  &  &  \\
\midrule
0.25 & \textbf{-3.57 ± 4.67} & 1.24 ± 3.66 & 0.75 ± 3.92 \\
0.75 & \textbf{-4.03 ± 5.39} & -0.94 ± 5.77 & -1.44 ± 5.52 \\
1 & \textbf{-4.36 ± 5.41} & -1.78 ± 5.76 & -1.81 ± 5.83 \\
1.5 & \textbf{-5.42 ± 5.48} & -2.83 ± 5.22 & -2.6 ± 5.85 \\
\bottomrule
\end{tabular}
}
\end{minipage}
\hfill
\begin{minipage}{0.48\textwidth}
\centering
\resizebox{\textwidth}{!}{\begin{tabular}{c|cccc}
\toprule
Learning Rate & 1e-03 & 1e-04 & 1e-05 \\
Arena Size (m) &  &  &  \\
\midrule
0.25 & -3.68 ± 4.57 & \textbf{2.12 ± 4.16} & \textbf{1.83 ± 3.8} \\
0.75 & -4.3 ± 5.18 & \textbf{1.42 ± 6.1} & \textbf{0.49 ± 5.26} \\
1 & -4.73 ± 5.29 & \textbf{0.13 ± 6.45} & \textbf{-0.1 ± 5.87} \\
1.5 & -5.72 ± 5.27 & \textbf{-1.13 ± 6.18} & \textbf{-0.66 ± 6.06} \\
\bottomrule
\end{tabular}
}
\end{minipage}

\begin{minipage}{0.48\textwidth}
\centering
\resizebox{\textwidth}{!}{\begin{tabular}{ccccc}
\toprule
Batch Size & 10 & 40 & 80 \\
Arena Size (m) &  &  &  \\
\midrule
0.25 & -3.54 ± 4.73 & -2.0 ± 4.58 & 3.96 ± 2.94 \\
0.75 & -5.73 ± 6.82 & -3.32 ± 6.09 & 2.62 ± 3.77 \\
1 & -6.32 ± 7.09 & -4.32 ± 6.16 & \textbf{2.7 ± 3.76} \\
1.5 & -8.5 ± 7.16 & -5.63 ± 5.93 & 3.28 ± 3.46 \\
\bottomrule
\end{tabular}
}
\end{minipage}
\hfill
\begin{minipage}{0.48\textwidth}
\centering
\resizebox{\textwidth}{!}{\begin{tabular}{c|cccc}
\toprule
Batch Size & 10 & 40 & 80 \\
Arena Size (m) &  &  &  \\
\midrule
0.25 & \textbf{-2.66 ± 4.79} & \textbf{-1.49 ± 4.81} & \textbf{4.43 ± 2.93} \\
0.75 & \textbf{-3.7 ± 6.65} & \textbf{-1.67 ± 6.12} & \textbf{2.97 ± 3.77} \\
1 & \textbf{-4.69 ± 7.29} & \textbf{-2.68 ± 6.33} & 2.67 ± 3.99 \\
1.5 & \textbf{-6.74 ± 7.4} & \textbf{-4.49 ± 6.52} & \textbf{3.72 ± 3.58} \\
\bottomrule
\end{tabular}
}
\end{minipage}

\begin{minipage}{0.48\textwidth}
\centering
\resizebox{\textwidth}{!}{\begin{tabular}{ccccc}
\toprule
Sequence Length & 40 & 80 & 120 \\
Arena Size (m) &  &  &  \\
\midrule
0.25 & -0.3 ± 4.05 & -0.35 ± 3.89 & -0.93 ± 4.31 \\
0.75 & -1.21 ± 5.34 & -2.04 ± 5.44 & -3.16 ± 5.9 \\
1 & -1.41 ± 5.19 & -2.9 ± 5.7 & -3.63 ± 6.11 \\
1.5 & -2.61 ± 5.36 & -3.21 ± 5.61 & -5.03 ± 5.58 \\
\bottomrule
\end{tabular}
}
\end{minipage}
\hfill
\begin{minipage}{0.48\textwidth}
\centering
\resizebox{\textwidth}{!}{\begin{tabular}{c|cccc}
\toprule
Sequence Length & 40 & 80 & 120 \\
Arena Size (m) &  &  &  \\
\midrule
0.25 & \textbf{0.1 ± 4.24} & \textbf{0.36 ± 4.03} & \textbf{-0.18 ± 4.26} \\
0.75 & \textbf{-0.09 ± 5.14} & \textbf{-0.99 ± 5.65} & \textbf{-1.32 ± 5.75} \\
1 & \textbf{-0.34 ± 5.5} & \textbf{-1.85 ± 5.95} & \textbf{-2.51 ± 6.16} \\
1.5 & \textbf{-1.63 ± 5.65} & \textbf{-2.38 ± 5.82} & \textbf{-3.5 ± 6.04} \\
\bottomrule
\end{tabular}
}
\end{minipage}

\begin{minipage}{0.48\textwidth}
\centering
\resizebox{\textwidth}{!}{\begin{tabular}{ccccc}
\toprule
Batch Size & 10 & 40 & 80 \\
Learning Rate &  &  &  \\
\midrule
1e-03 & -5.2 ± 6.95 & -2.81 ± 5.8 & 4.77 ± 2.55 \\
1e-04 & -5.07 ± 6.71 & -2.56 ± 5.66 & 3.81 ± 3.47 \\
1e-05 & \textbf{-7.79 ± 5.69} & \textbf{-6.08 ± 5.61} & \textbf{0.84 ± 4.42} \\
\bottomrule
\end{tabular}
}
\end{minipage}
\hfill
\begin{minipage}{0.48\textwidth}
\centering
\resizebox{\textwidth}{!}{\begin{tabular}{c|cccc}
\toprule
Batch Size & 10 & 40 & 80 \\
Learning Rate &  &  &  \\
\midrule
1e-03 & \textbf{-2.43 ± 7.63} & \textbf{-0.68 ± 6.56} & \textbf{5.02 ± 2.97} \\
1e-04 & \textbf{-2.73 ± 6.64} & \textbf{-0.76 ± 5.8} & \textbf{4.65 ± 3.29} \\
1e-05 & -8.18 ± 5.33 & -6.31 ± 5.46 & 0.67 ± 4.44 \\
\bottomrule
\end{tabular}
}
\end{minipage}

\begin{minipage}{0.48\textwidth}
\centering
\resizebox{\textwidth}{!}{\begin{tabular}{ccccc}
\toprule
Sequence Length & 40 & 80 & 120 \\
Learning Rate &  &  &  \\
\midrule
1e-03 & -0.76 ± 5.25 & -0.82 ± 4.9 & -1.66 ± 5.15 \\
1e-04 & -0.77 ± 4.98 & -1.18 ± 5.29 & -1.88 ± 5.57 \\
1e-05 & \textbf{-2.62 ± 4.72} & \textbf{-4.38 ± 5.29} & \textbf{-6.03 ± 5.7} \\
\bottomrule
\end{tabular}
}
\end{minipage}
\hfill
\begin{minipage}{0.48\textwidth}
\centering
\resizebox{\textwidth}{!}{\begin{tabular}{c|cccc}
\toprule
Sequence Length & 40 & 80 & 120 \\
Learning Rate &  &  &  \\
\midrule
1e-03 & \textbf{0.86 ± 5.75} & \textbf{0.74 ± 5.54} & \textbf{0.3 ± 5.88} \\
1e-04 & \textbf{0.58 ± 5.14} & \textbf{0.31 ± 5.38} & \textbf{0.28 ± 5.21} \\
1e-05 & -2.92 ± 4.5 & -4.69 ± 5.17 & -6.21 ± 5.56 \\
\bottomrule
\end{tabular}
}
\end{minipage}

\begin{minipage}{0.48\textwidth}
\centering
\resizebox{\textwidth}{!}{\begin{tabular}{ccccc}
\toprule
Sequence Length & 40 & 80 & 120 \\
Batch Size &  &  &  \\
\midrule
10 & 4.31 ± 2.83 & 3.45 ± 3.43 & 1.65 ± 4.18 \\
40 & -3.09 ± 5.55 & -3.9 ± 5.78 & -4.46 ± 5.74 \\
80 & -5.38 ± 6.58 & -5.92 ± 6.27 & -6.77 ± 6.51 \\
\bottomrule
\end{tabular}
}
\end{minipage}
\hfill
\begin{minipage}{0.48\textwidth}
\centering
\resizebox{\textwidth}{!}{\begin{tabular}{c|cccc}
\toprule
Sequence Length & 40 & 80 & 120 \\
Batch Size &  &  &  \\
\midrule
10 & \textbf{4.47 ± 2.9} & \textbf{3.56 ± 3.6} & \textbf{2.31 ± 4.21} \\
40 & \textbf{-1.97 ± 5.85} & \textbf{-2.86 ± 5.98} & \textbf{-2.92 ± 6.0} \\
80 & \textbf{-3.98 ± 6.65} & \textbf{-4.35 ± 6.51} & \textbf{-5.02 ± 6.44} \\
\bottomrule
\end{tabular}
}
\end{minipage}

\mycomment{
\bigskip

We perform a similar analysis for 64 place cells, training three models for each combination of the following parameters. The mean place cell score ($\pm$ standard deviation) is presented below across various pairs of parameters.

\begin{itemize}
    \item Arena Size (in meters): 0.25, 0.75, 1, 1.5, 3
    \item Number of Hidden Units (Ng): 256, 512, 1028, 2046
    \item Learning Rate: 0.0001
    \item Batch Size: 10, 40, 80
    \item Sequence Length: 200
\end{itemize}

\begin{minipage}{0.48\textwidth}
\centering
    \textbf{Skaggs PC Score} ($N_p  = 64$)
\end{minipage}
\hfill
\begin{minipage}{0.48\textwidth}
\centering
    \textbf{Spectral PC Score} ($N_p  = 64$)
\end{minipage}
\hfill
\begin{minipage}{0.48\textwidth}
\centering
\resizebox{\textwidth}{!}{\begin{tabular}{cccccc}
\toprule
Ng & 256 & 512 & 1028 & 2048 \\
\midrule
Arena Size (m) & & & & \\
0.25 & -6.38 ± 5.09 & \textbf{-3.12 ± 5.61} & \textbf{-2.32 ± 5.21} & -3.32 ± 4.44 \\
0.75 & \textbf{-14.24 ± 6.98} & \textbf{-11.77 ± 9.4} & \textbf{-12.21 ± 10.17} & -15.17 ± 7.68 \\
1 & -10.93 ± 7.1 & \textbf{-7.66 ± 9.11} & -8.89 ± 9.36 & -11.51 ± 8.58 \\
1.5 & -9.22 ± 7.26 & -5.49 ± 8.03 & -4.32 ± 7.94 & -7.01 ± 8.33 \\
3 & -9.13 ± 7.67 & -6.46 ± 8.1 & -5.97 ± 8.58 & -8.27 ± 8.66 \\
\bottomrule
\end{tabular}
}
\end{minipage}
\hfill
\begin{minipage}{0.48\textwidth}
\centering
\resizebox{\textwidth}{!}{\begin{tabular}{c|cccc}
\toprule
Ng & 256 & 512 & 1028 & 2048 \\
\midrule
Arena Size (m) & & & & \\
0.25 & \textbf{-5.83 ± 4.08} & -3.99 ± 3.8 & -3.71 ± 3.58 & \textbf{-2.78 ± 4.24} \\
0.75 & -15.49 ± 4.31 & -12.41 ± 7.08 & -13.17 ± 8.64 & \textbf{-13.69 ± 8.03} \\
1 & \textbf{-9.39 ± 6.42} & -8.26 ± 6.86 & \textbf{-8.16 ± 7.81} & \textbf{-9.24 ± 8.39} \\
1.5 & \textbf{-6.34 ± 6.68} & \textbf{-4.65 ± 6.46} & \textbf{-4.2 ± 6.74} & \textbf{-4.63 ± 6.34} \\
3 & \textbf{-7.19 ± 7.03} & \textbf{-5.8 ± 6.36} & \textbf{-5.46 ± 7.2} & \textbf{-6.32 ± 7.24} \\
\bottomrule
\end{tabular}
}
\end{minipage}

\begin{minipage}{0.48\textwidth}
\centering
\resizebox{\textwidth}{!}{\begin{tabular}{ccccc}
\toprule
Learning Rate & 1e-04 \\
Ng &  \\
\midrule
256 & -9.98 ± 6.82 \\
512 & \textbf{-6.9 ± 8.05} \\
1028 & \textbf{-6.74 ± 8.25} \\
2048 & -9.05 ± 7.54 \\
\bottomrule
\end{tabular}
}
\end{minipage}
\hfill
\begin{minipage}{0.48\textwidth}
\centering
\resizebox{\textwidth}{!}{\begin{tabular}{c|cccc}
\toprule
Learning Rate & 1e-04 \\
Ng &  \\
\midrule
256 & \textbf{-8.85 ± 5.7} \\
512 & -7.02 ± 6.11 \\
1028 & -6.94 ± 6.8 \\
2048 & \textbf{-7.33 ± 6.85} \\
\bottomrule
\end{tabular}
}
\end{minipage}

\begin{minipage}{0.48\textwidth}
\centering
\resizebox{\textwidth}{!}{\begin{tabular}{ccccc}
\toprule
Batch Size & 10 & 40 & 80 \\
Ng &  &  &  \\
\midrule
256 & -10.14 ± 6.9 & \textbf{-9.46 ± 6.35} & -10.33 ± 7.21 \\
512 & \textbf{-7.52 ± 8.32} & \textbf{-7.15 ± 7.85} & -6.03 ± 7.97 \\
1028 & -7.97 ± 9.03 & \textbf{-7.56 ± 8.3} & -4.69 ± 7.43 \\
2048 & -10.4 ± 8.13 & -9.89 ± 7.93 & -6.88 ± 6.55 \\
\bottomrule
\end{tabular}
}
\end{minipage}
\hfill
\begin{minipage}{0.48\textwidth}
\centering
\resizebox{\textwidth}{!}{\begin{tabular}{c|cccc}
\toprule
Batch Size & 10 & 40 & 80 \\
Ng &  &  &  \\
\midrule
256 & \textbf{-9.23 ± 5.37} & -9.64 ± 5.19 & \textbf{-7.67 ± 6.56} \\
512 & -7.8 ± 6.33 & -7.87 ± 5.72 & \textbf{-5.4 ± 6.28} \\
1028 & \textbf{-7.96 ± 6.92} & -8.62 ± 6.64 & \textbf{-4.23 ± 6.82} \\
2048 & \textbf{-8.7 ± 7.09} & \textbf{-8.92 ± 6.63} & \textbf{-4.37 ± 6.82} \\
\bottomrule
\end{tabular}
}
\end{minipage}

\begin{minipage}{0.48\textwidth}
\centering
\resizebox{\textwidth}{!}{\begin{tabular}{ccccc}
\toprule
Sequence Length & 200 \\
Ng &  \\
\midrule
256 & -9.98 ± 6.82 \\
512 & \textbf{-6.9 ± 8.05} \\
1028 & \textbf{-6.74 ± 8.25} \\
2048 & -9.05 ± 7.54 \\
\bottomrule
\end{tabular}
}
\end{minipage}
\hfill
\begin{minipage}{0.48\textwidth}
\centering
\resizebox{\textwidth}{!}{\begin{tabular}{c|cccc}
\toprule
Sequence Length & 200 \\
Ng &  \\
\midrule
256 & \textbf{-8.85 ± 5.7} \\
512 & -7.02 ± 6.11 \\
1028 & -6.94 ± 6.8 \\
2048 & \textbf{-7.33 ± 6.85} \\
\bottomrule
\end{tabular}
}
\end{minipage}

\begin{minipage}{0.48\textwidth}
\centering
\resizebox{\textwidth}{!}{\begin{tabular}{ccccc}
\toprule
Learning Rate & 1e-04 \\
Arena Size (m) &  \\
\midrule
0.25 & \textbf{-3.78 ± 5.09} \\
0.75 & -6.51 ± 7.89 \\
1 & -7.46 ± 8.25 \\
1.5 & -9.75 ± 8.54 \\
3 & \textbf{-13.35 ± 8.56} \\
\bottomrule
\end{tabular}
}
\end{minipage}
\hfill
\begin{minipage}{0.48\textwidth}
\centering
\resizebox{\textwidth}{!}{\begin{tabular}{c|cccc}
\toprule
Learning Rate & 1e-04 \\
Arena Size (m) &  \\
\midrule
0.25 & -4.08 ± 3.93 \\
0.75 & \textbf{-4.95 ± 6.55} \\
1 & \textbf{-6.19 ± 6.96} \\
1.5 & \textbf{-8.76 ± 7.37} \\
3 & -13.69 ± 7.02 \\
\bottomrule
\end{tabular}
}
\end{minipage}

\begin{minipage}{0.48\textwidth}
\centering
\resizebox{\textwidth}{!}{\begin{tabular}{ccccc}
\toprule
Batch Size & 10 & 40 & 80 \\
Arena Size (m) &  &  &  \\
\midrule
0.25 & \textbf{-4.36 ± 5.13} & \textbf{-3.68 ± 4.69} & -3.31 ± 5.44 \\
0.75 & -7.03 ± 7.95 & -6.85 ± 7.85 & -5.64 ± 7.87 \\
1 & -7.72 ± 9.07 & -8.08 ± 7.99 & -6.57 ± 7.69 \\
1.5 & -11.26 ± 9.17 & \textbf{-10.21 ± 8.78} & -7.77 ± 7.66 \\
3 & -14.68 ± 9.16 & \textbf{-13.75 ± 8.72} & -11.61 ± 7.79 \\
\bottomrule
\end{tabular}
}
\end{minipage}
\hfill
\begin{minipage}{0.48\textwidth}
\centering
\resizebox{\textwidth}{!}{\begin{tabular}{c|cccc}
\toprule
Batch Size & 10 & 40 & 80 \\
Arena Size (m) &  &  &  \\
\midrule
0.25 & -5.44 ± 3.75 & -4.31 ± 3.58 & \textbf{-2.49 ± 4.46} \\
0.75 & \textbf{-5.39 ± 6.52} & \textbf{-6.33 ± 6.3} & \textbf{-3.14 ± 6.85} \\
1 & \textbf{-6.82 ± 7.14} & \textbf{-7.18 ± 6.59} & \textbf{-4.58 ± 7.14} \\
1.5 & \textbf{-10.02 ± 7.39} & -10.43 ± 7.33 & \textbf{-5.84 ± 7.39} \\
3 & \textbf{-14.44 ± 7.36} & -15.58 ± 6.42 & \textbf{-11.05 ± 7.28} \\
\bottomrule
\end{tabular}
}
\end{minipage}

\begin{minipage}{0.48\textwidth}
\centering
\resizebox{\textwidth}{!}{\begin{tabular}{ccccc}
\toprule
Sequence Length & 200 \\
Arena Size (m) &  \\
\midrule
0.25 & \textbf{-3.78 ± 5.09} \\
0.75 & -6.51 ± 7.89 \\
1 & -7.46 ± 8.25 \\
1.5 & -9.75 ± 8.54 \\
3 & \textbf{-13.35 ± 8.56} \\
\bottomrule
\end{tabular}
}
\end{minipage}
\hfill
\begin{minipage}{0.48\textwidth}
\centering
\resizebox{\textwidth}{!}{\begin{tabular}{c|cccc}
\toprule
Sequence Length & 200 \\
Arena Size (m) &  \\
\midrule
0.25 & -4.08 ± 3.93 \\
0.75 & \textbf{-4.95 ± 6.55} \\
1 & \textbf{-6.19 ± 6.96} \\
1.5 & \textbf{-8.76 ± 7.37} \\
3 & -13.69 ± 7.02 \\
\bottomrule
\end{tabular}
}
\end{minipage}

\begin{minipage}{0.48\textwidth}
\centering
\resizebox{\textwidth}{!}{\begin{tabular}{ccccc}
\toprule
Batch Size & 10 & 40 & 80 \\
Learning Rate &  &  &  \\
\midrule
1e-04 & -9.01 ± 8.1 & \textbf{-8.52 ± 7.61} & -6.98 ± 7.29 \\
\bottomrule
\end{tabular}
}
\end{minipage}
\hfill
\begin{minipage}{0.48\textwidth}
\centering
\resizebox{\textwidth}{!}{\begin{tabular}{c|cccc}
\toprule
Batch Size & 10 & 40 & 80 \\
Learning Rate &  &  &  \\
\midrule
1e-04 & \textbf{-8.42 ± 6.43} & -8.76 ± 6.04 & \textbf{-5.42 ± 6.62} \\
\bottomrule
\end{tabular}
}
\end{minipage}

\begin{minipage}{0.48\textwidth}
\centering
\resizebox{\textwidth}{!}{\begin{tabular}{ccccc}
\toprule
Sequence Length & 200 \\
Learning Rate &  \\
\midrule
1e-04 & -8.17 ± 7.66 \\
\bottomrule
\end{tabular}
}
\end{minipage}
\hfill
\begin{minipage}{0.48\textwidth}
\centering
\resizebox{\textwidth}{!}{\begin{tabular}{c|cccc}
\toprule
Sequence Length & 200 \\
Learning Rate &  \\
\midrule
1e-04 & \textbf{-7.54 ± 6.37} \\
\bottomrule
\end{tabular}
}
\end{minipage}

\begin{minipage}{0.48\textwidth}
\centering
\resizebox{\textwidth}{!}{\begin{tabular}{ccccc}
\toprule
Sequence Length & 200 \\
Batch Size &  \\
\midrule
10 & -6.98 ± 7.29 \\
40 & -9.01 ± 8.1 \\
80 & \textbf{-8.52 ± 7.61} \\
\bottomrule
\end{tabular}
}
\end{minipage}
\hfill
\begin{minipage}{0.48\textwidth}
\centering
\resizebox{\textwidth}{!}{\begin{tabular}{c|cccc}
\toprule
Sequence Length & 200 \\
Batch Size &  \\
\midrule
10 & \textbf{-5.42 ± 6.62} \\
40 & \textbf{-8.42 ± 6.43} \\
80 & -8.76 ± 6.04 \\
\bottomrule
\end{tabular}
}
\end{minipage}
}

\end{document}